\documentclass[aps,pre,reprint,twoside,longbibliography,groupedaddress]{revtex4-1}
\usepackage{amsmath,amsfonts,amssymb,amsthm,graphicx,color,epstopdf,tipa}
\usepackage{natbib}

\newcommand{\lu}{\textupsilon}
\newcommand{\su}{\textturnv}

\newcommand{\dt}{\delta t}
\newcommand{\Dx}{\Delta x}
\newcommand{\Dy}{\Delta y}
\newcommand{\sgn}{\text{sgn}}

\newcommand{\ra}{\rightarrow}

\newcommand{\EE}{ \mathbf{E} }
\newcommand{\p}{\partial}
\newcommand{\unit}[1]{\hat{\mathbf{#1}}}
\newcommand{\bv}[1]{\boldsymbol{\mathbf{#1}}}

\begin{document}


\title{Spatial evolution of human dialects}


\author{James Burridge}
\email[]{james.burridge@port.ac.uk}
\affiliation{Department of Mathematics, University of Portsmouth}


\date{\today}

\begin{abstract}

The geographical pattern of human dialects is a result of history.  Here, we formulate a simple spatial model of language change which shows that the final result of this historical evolution may, to some extent, be predictable. The model shows that the boundaries of language dialect regions are controlled by a length minimizing effect analogous to surface tension, mediated by variations in population density which can induce curvature, and by the shape of coastline or similar borders. The predictability of dialect regions arises because these effects will drive many complex, randomized early states toward one of a smaller number of stable final configurations. The model is able to reproduce observations and predictions of dialectologists. These include dialect continua, isogloss bundling, fanning, the wave-like spread of dialect features from cities, and the impact of human movement on the number of dialects that an area can support. The model also provides an analytical form for S\'{e}guy's Curve giving the relationship between geographical and linguistic distance, and a generalisation of the curve to account for the presence of a population centre. A simple modification allows us to analytically characterize the variation of language use by age in an area undergoing linguistic change.
\end{abstract}


\maketitle

\section{Introduction \label{intro}}

	Over time, human societies develop systems of belief, languages, technology and artistic forms which collectively may be called \textit{culture}. The formation of culture requires individuals to have ideas, and then for others to copy them. Historically, most copying has required face-to-face interaction, and because most human beings tend to remain localized in geographical regions which are small in comparison to the world, then human culture can take quite different forms in different places. One aspect of culture where geographical distribution has been studied in great detail is dialect \cite{cha98}.

In order to visualize the spatial extent of dialects, dialectologists have traditionally drawn \textit{Isoglosses}: lines enclosing the domain within which a particular linguistic feature (a word, a phoneme or an element of syntax) is used. However, it is not usually the case that language use changes abruptly at an isogloss - typically there is a transition zone where a mixture of alternative features is used \cite{cha98}. In fact, there is debate about whether the most appropriate way to view the geographical organization of dialects is as a set of distinct areas, or as a continuum without sharp boundaries \cite{hee01, cha98, blo33}. Whereas an isogloss represents the extent of an individual feature, a recognisable dialect is typically a combination of many distinctive features \cite{cha98,tru99}. We can attempt to distinguish dialects by superposing many different isoglosses,  but often they do not coincide \cite{blo33} leading to ambiguous conclusions.

The first steps toward an objective, quantitative analysis of the shapes of dialect areas were made by S\'{e}guy \cite{seg71,seg73}, who examined large aggregates of features, making comparison between lexical distances and geographic separations. Central to the quantitative study of dialects, called \textit{dialectometry} (see \cite{wie15} for a recent review), is the measurement of linguistic distance which, for example, can be viewed as the smallest number of insertions, deletions or substitutions of language features needed to transform one segment of speech into another \cite{kes95}. This ``Levenshtein distance'' was originally devised to measure the difference between sequences \cite{lev66}.  Using a metric of this kind, a set of dialect observations can be grouped into clusters according to their linguistic (as opposed to spatial) closeness \cite{sha07,emb13,wie11,ner11}. The clusters then define geographical dialect areas. 

The question we address is \textit{why} dialect domains have particular spatial forms, and to give a quantitative answer requires a model. The question has been addressed in the past, famously (amongst dialectologists) by Trudgill \cite{tru74,cha98}, with his ``gravity model''. According to this, the strength of linguistic interaction between two population centres is proportional to the product of their populations, divided by the square of the distance between them. The \textit{influence} of a settlement (e.g. a city) $i$ on another, $j$, is then defined to be the product of interaction strength with the ratio $P_i/(P_i+P_j)$ where $P_i$ and $P_j$ are the population sizes of settlements $i$ and $j$.  These additive influence scores may then be used to predict the progress of a linguistic change which originated in one city, by determining the settlements over which it exerts greatest net influence. It is then predicted that the change progresses from settlement to settlement in a \textit{cascade}. Predictions may also be made regarding the combined influence of cities on neighbouring non-urban areas. The model has been partially successful in predicting observed sequences of linguistic change \cite{cha98,wol03,lab03,bai93}, and offers some qualitative insight into the most likely positions of isoglosses \cite{tru74}. In this paper we offer an alternative model, also based on population data, which makes use of ideas from Statistical Mechanics. Rather than starting with a postulate about the nature of interactions between population centres, we begin with assumptions about the interactions between speakers. From these assumptions about small scale behaviour we derive predictions about macroscopic behaviour. This approach has the advantage of making clear the link between individual human interactions, and population level behaviour. Moreover we are able to unambiguously define the dynamics of the model, and make precise predictions about the locations of isoglosses, the nature of transition regions between linguistic forms, and the most likely structure of dialect domains. There are links between our approach and agent based models of language change \cite{sta13}, which directly simulate the behaviour of individuals. The difference between this approach and ours lies in the fact that for us,  assumptions about individual behaviour lead to equations for language evolution which are macroscopic in character. These equations have considerable analytical tractability and offer a simple and intuitive picture of the large scale spatial processes at play. 

In seeking to model the spatial distribution of language beginning with the individual, we are encouraged by the fact that dialects are created through a vast number of complex interactions between millions of people. These people are analogous to atoms in the physical context, and when very large numbers of particles interact in physical systems, simple macroscopic laws often emerge. Despite the fact dialects are the product of hundreds of years of linguistic and cultural evolution \cite{tru99}, so historical events must have played a role in creating their spatial distribution \cite{cry03}, the physical analogy suggests that it may be possible to formulate approximate statistical laws which play a powerful role in their spatial evolution.

A physical effect analogous to the formation of dialects is \textit{phase ordering} \cite{bra94}. This occurs, for example, in ferromagnetic materials, where each atom attempts to align itself with neighbours. If the material is two dimensional (a flat sheet) this leads to the formation of a patchwork of domains where all atoms are aligned with others in the same domain, but not with those in other domains.  The boundaries between these regions of aligned atoms evolve so as to minimize boundary length \cite{bar09,kra10}. The human agents who interact to form dialects behave in roughly the same way (as do some birds \cite{bur16}). 	When people speak and listen to each other, they have a tendency to conform to the patterns of speech they hear others using, and therefore to `align' their dialects. Since people typically remain geographically localized in their everyday lives, they tend to align with those nearby. This local copying gives rise to dialects in the same way that short range atomic interactions give rise to domains in ferromagnets. However, whereas the atoms in a ferromagnet are regularly spaced, human population density is variable. We will show that as a result, stable boundaries between domains become curved lines. 

While our interest is in  the spatial distribution of linguistic forms, there are other properties of language for which parallels with the physical or natural world can be usefully drawn, and corresponding mathematical methods applied. For example the rank-frequency distribution of word use, compiled from millions of books, takes the form of a double power law \cite{ger13,pet12}, which can be explained \cite{ger13} using a novel form of the Yule process \cite{wil22,new05}, first introduced to explain the distribution of the number of species in genera of flowering plants. Historical fluctuations in the relative frequency with which words are used have been shown to decay as a language ages and expands \cite{pet12}, analogous with the cooling effect produced by the expansion of a gas. Methods used to understand disorder in physical systems (``quenched'' averages), have been applied to explain how a tendency to focus on topics controls fluctuations in the combined vocabulary of groups of texts \cite{ger14}. A significant focus of current Statistical Physics research has been on the evolution and properties of networks \cite{new10}, which have many diverse applications from the spread of ideas, fashions and disease \cite{gle16}, to the vulnerability of the internet \cite{alb00}. Real networks are often formed by ``preferential attachment'' where new connections are more often made to already well connected nodes, leading to a ``scale free'' (power law) distribution of node degree. The popularity of words has been shown to evolve in the same way \cite{per12}; words used more in the past tend to be used more in the future. Beyond the study of word use and vocabulary, agent based models such as \textit{the naming game} \cite{lor11},  used to investigate the emergence of language, and the \textit{utterance selection model} \cite{bax06}, used to model changes in language use over time have been particularly influential. We follow the latter model by representing language use using a set of discrete \textit{linguistic variables}. Spatial models motivated by concepts of statistical physics have also been used to study the spread of crime \cite{ors15} and to devise optimal vaccination strategies to prevent disease \cite{wan16}. The importance of the emergence of order in social contexts, and connections to Statistical Physics, may be found in a wide ranging review \cite{Cas09} by Castellano et al. 


\section{Summary for Linguists}

\subsection{Contents of the paper}
The aim of this paper is to adapt the theory of phase ordering to the study of dialects, and then to use this theory to explain aspects of their spatial structure. For those without a particular mathematical or quantitative inclination, the model can be simply explained: We assume that people come into linguistic contact predominantly with those who live within a typical travel radius of their home (around 10 to 20 km). If they live near a town or city, we assume that they experience more frequent interactions with people from the city than with those living outside it, simply because there are many more city dwellers with whom to interact. We represent dialects using a set of \textit{linguistic variables} \cite{cha98}, and we suppose that speakers have a tendency to adapt their speech over time in order to conform to local conventions of language use. Our model is deliberately minimal: these are our \textit{only} assumptions. We discover that, starting from any historical language state, these assumptions lead to the formation of spatial domains where particular linguistic variants are in common use, as in Figure \ref{fig:evo}. We find that the isoglosses which bound these domains are driven away from population centres, that they tend to reduce in curvature over time, and that they are most stable when emerging perpendicular to borders of a linguistic domain. These theoretical \textit{principles of isogloss evolution} are explained pictorially in Figures \ref{fig:tension}, \ref{fig:repulsion} and \ref{fig:coast}, and provide a theoretical explanation for a range of observed phenomena, such as the dialects of England (Figure \ref{fig:clust}), the Rhenish Fan (Figure \ref{fig:fan}),  the wave-like spread of language features from cities (Figures \ref{fig:city} and \ref{fig:segCity}), the fact that narrow regions often have ``striped'' dialects (Figure \ref{fig:stripe}), and that coastal indentations including rivers and estuaries often generate isogloss bundles. Our assumptions also lead to a mathematical expression for the relationship between linguistic and geographical distance -- the ``S\'{e}guy Curve'' -- and a hypothesis regarding question of when dialects should be viewed as a spatial continuum, as opposed to distinct areas (Figure \ref{fig:nations}).

\subsection{How might a linguist make use of this work?}
Without using mathematics, but having understood our principles of isogloss evolution and considered the examples set out in this paper, further cases may be sought where the principles  explain observations. If the principles cannot explain a particular situation or are violated, one might seek to understand what was missing from the underlying assumptions, or if they were wrong. Since the assumptions are so minimal, they cannot be the whole story, and a discussion of possible missing pieces is given in the conclusion. For the mathematically inclined linguist, appendix \ref{discrete} sets out an elementary scheme for solving the fundamental evolution equation on a computer. This scheme also offers a simple and intuitive understanding of the model, and can be implemented using only a spreadsheet (see supplementary material), although a computer program would be much faster. Using this, isogloss evolution can be explored in linguistic domains with any shape and population distribution. The simplicity of the scheme  invites adaptation to include more linguistic realism (e.g. bias toward a linguistic variant). Beyond the exploration of individual isoglosses, a line of inquiry which may be of interest to dialectometrists is to test our predicted forms of S\'{e}guy's Curve against observations.


\section{The model}

Our aim is to define a model of speech copying which incorporates as few assumptions as possible, whilst allowing the effect of local linguistic interaction and movement to be investigated. The model has its roots in the ideas of the linguist Leonard Bloomfield \cite{blo33} who believed that the speech pattern of an individual constantly evolved through his or her life via pairwise interaction. This microscopic view of language change lead to the prediction that the diffusion of linguistic features should follow routes with the greatest \textit{density of communication}. Bloomfield defined this as the density of conversational links between speakers accumulated over a given period of time. In our model the analogy of this link density is an interaction kernel weighted by spatial variations in population distribution. We implicitly assume that interaction is inherently local so that linguistic changes spread via \textit{normal contact} \cite{ner10}, rather than via major displacements, conquests, or dispersion of settled communities. We are therefore modelling language in stable settlements, with initial conditions set by the most recent major population upheaval.

We consider a population of speakers, each of whom has a small home neighbourhood, and we introduce a population density, $\rho(x,y)$, giving the spatial variation of the number homes per unit area. 
In order to incorporate local human movement within the model, we begin by defining a Gaussian  
\textit{interaction kernel} for each speaker 
\begin{align*}
\phi(\Dx,\Dy) := \frac{1}{2\pi \sigma^2} \exp \left\{ -\frac{\Dx^2+\Dy^2}{2 \sigma^2} \right\}.
\end{align*}
Note that the symbol $:=$ indicates the \textit{definition} of a new quantity. Consider a speaker, \textit{Anna}, whose home neighbourhood is centred on $(x_0,y_0)$. In the absence of variation in population density, $\phi$ is the normalized distribution of the relative positions, $(\Dx, \Dy)$, of the home neighbourhoods of speakers with whom Anna regularly interacts. The constant $\sigma$, the \textit{interaction range}, is a measure of the typical  geographical distance between the neighbourhoods of interacting speakers. Now suppose that density is not uniform due to the presence of a city, or a sparsely populated mountainous area. In this case while Anna is going about her daily life she is more likely to hold conversations with people whose homes lie in a nearby densely populated region because these people constitute a greater proportion of the local population. To incorporate this density effect we define a normalized \textit{weighted interaction kernel} for a home at $(x_0,y_0)$
\begin{align*}
k(x_0,y_0;x,y): = \frac{\phi(x-x_0,y-y_0) \rho(x,y)}{\int_{\mathbb{R}^2} \phi(u-x_0,v-y_0) \rho(u,v) du dv}.
\end{align*}
Given any region $\mathcal{A}$, the fraction of Anna's interactions that are with people who live in $\mathcal{A}$ is $\int_{\mathcal{A}} k(x_0,y_0,x,y) dx dy$.

We distinguish between dialects by constructing a set of \textit{linguistic variables} whose values vary between dialects.  A single variable might, for example, be the pronunciation of the vowel \textit{u} in the words `but' and `up' \cite{tru99}.  In England, northerners use a long form:  `boott' and `oopp', with phonetic symbol  [\lu], and southerners use a short version, [\su] .  Considering	a single variable which we suppose has $V>1$ variants, we define $f_i(x,y,t)$ to be the relative frequency with which the $i$th variant of our variable is used by speakers in the neighbourhood of $(x,y)$, at time $t$. For mathematical simplicity we assume that nearby speakers use language in a similar way, so that $f_i(x,y,t)$ varies smoothly with position.

People speak on average 16,000 words per day \cite{meh07} and can take months or years (depending on their age and background) to adapt their speech to local forms \cite{cha92, sie10}.  Changing speech habits therefore involves a very large number of word exchanges, at least in the tens of thousands (comparable in magnitude to typical vocabulary size \cite{blo00}). Although the rate at which individuals adapt their speech is not constant throughout life (it is particularly rapid in the young), adaptation has been observed even in late middle age \cite{san07}. 
To capture the cumulative effect of linguistic interaction
we make use of a \textit{forgetting curve}, which measures the relative importance of recent interactions to older ones. From a mathematical point of view, the simplest form for this curve is an exponential, and in fact there is some evidence from experiments involving word recall \cite{ave10} which suggests that this is an appropriate choice. However, we emphasize that the curve, for us, is simply a way to capture the fact that  current speech patterns depend on past interactions and that older interactions tend to be less important. With this in mind we make the following definition of the memory of a speaker from the neighbourhood of $(x,y)$, for the $i$th variant of a variable
\begin{align}
m_i(x,y,t) & := \nonumber   \\
\int_{-\infty}^{t} & \frac{e^{\frac{s-t}{\tau}}}{\tau}  \left[ \int_{\mathbb{R}^2} k(x,y;u,v)f_i(u,v,s) du dv \right] ds \label{eq:mint} \\
 \approx  \int_{-\infty}^{t} & \frac{e^{\frac{s-t}{\tau}}}{\tau}  \Big[ f_i(x,y,s) \nonumber   \\ 
& +  \frac{\sigma^2}{2 \rho(x,y)} \nabla^2 \{\rho(x,y) f_i(x,y,s) \}\Big]ds. \label{eq:m}
\end{align} 
An intuitive understanding of this equation may be gained by imagining that each speaker possesses an internal tape recorder which records language use as they travel around the vicinity of their home. As time passes, older recordings fade in importance to the speaker, and the variable $m_i$ measures the historical frequency with which variable $i$ has been heard, accounting for the declining importance of older recordings. The rate of this decline is determined by the parameter $\tau$, which we call \textit{memory length}, and note that changing its value simply rescales the unit of time. We note also that this form of memory may be seen as a deterministic spatial version of the discrete stochastic memory used in the Utterance Selection Model \cite{bly12,bax06}. On the grounds that speakers collect very large samples of local linguistic information, our definition does not contain terms representing random sampling error. In going from equation (\ref{eq:mint}) to (\ref{eq:m}) we have used the saddle point method \cite{ma71} to approximate the spatial integral in equation (\ref{eq:mint}) and assumed that $|\nabla^2 \rho|/\rho$ is small compared to $\sigma^2$ (that is, population changes approximately linearly over the length scale of human interaction). 

To allow speakers to base their current speech on what they have heard in the past we let $f_i(x,y,t)$ be a function, $p_i$, of the set of memories $(m_1,m_2,\ldots,m_V)=:\bv{m}$ 
\begin{align*}
f_i(x,y,t) := p_i[\bv{m}(x,y,t)].
\end{align*}
Differentiating equation (\ref{eq:m}) with respect to $t$, and rescaling the units of time so that one time unit is equal to one memory length $\tau$, we obtain 
\begin{multline}
\frac{\p m_i(x,y,t)}{\p t} = p_i[\bv{m}(x,y,t)] - m_i(x,y,t) \\ 
+\frac{\sigma^2}{2 \rho(x,y)} \nabla^2 \{\rho(x,y) p_i[\bv{m}(x,y,t)] \},
\label{evo}
\end{multline}
which governs the spatial evolution of the $i$th alternative for a single linguistic variable. We note that memory length no longer appears as a parameter. An enhanced intuitive understanding of this evolution equation may be gained from its discrete counterpart, used to find computational solutions, and derived in appendix \ref{discrete}.

The simplest possible choice for $p_i$ is to let speakers use each variant with the same frequency that they remember it being used: $p_i[\bv{m}(x,y,t)] = m_i(x,y,t)$. This produces ``neutral evolution'' \cite{bly12,bax06,cro00,kim83,mor58} where there is no bias in the evolution of each variant. Equation (\ref{evo}) then describes pure diffusion, and variants spread out uniformly over the system. If all linguistic variables evolved in this way we would eventually have one spatially homogeneous mixture of grammar, pronunciation  and vocabulary. If our memory model involved a stochastic component \cite{bly12} then eventually we would expect all but one variant of each variable to disappear. Neither of these outcomes reflects the reality of locally distinctive forms of language. 

\begin{figure}
	\centering
	\includegraphics[width=.95\linewidth]{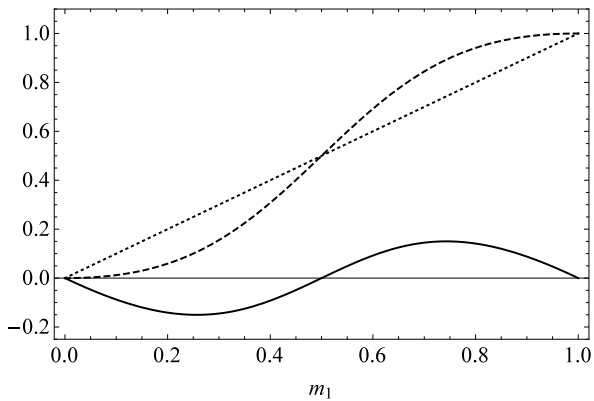}
	\caption{ Dashed line shows the function $p_1(\bv{m}) \equiv f_1$, defined by equation (\ref{eqn:pk}) for the $V=2$ model with conformity number $\beta=2$. Dotted line shows the neutral version $p_1(\bv{m})=m_1$ (when $\beta=1$) for comparison. Solid line shows the function $p_1(\bv{m})-m_1$ in the case $\beta=2$, giving the time derivative of the memory in the absence of spatial variation. Note, dashed line shows that when speakers' memories contain a majority of variant 1, then they use this variant in a greater proportion then they recall it being used. This leads to progressively greater levels of conformity: more speakers using variant 1. }
	\label{fig:pFunc}
\end{figure}

We motivate our choice for $p_i$ based on two observations. The first is that dialects exist. In order for this to be the case, if the $i$th variant of a linguistic variable has been established amongst a local population for a considerable time so that $m_i \approx 1$, then a small amount of immigration into the region by speakers using a different variable should not normally be sufficient to change it. Mathematically, this is equivalent to the statement that the non diffusive term, $p_i[\bv{m}(x,y,t)] - m_i(x,y,t)$, in our evolution equation (\ref{evo}) must possess a locally stable fixed point at $m_i=1$. The second observation is derived from experiments on social learning, which show that the behaviour of individuals is considerably influenced by the majority opinion of those with whom they interact \cite{asc56,bon96,mor12}. In fact, such social conformity is widely observed in the animal kingdom and is responsible for the formation of dialects in some species of birds \cite{Nel00}. Recent experimental research into human social learning \cite{mor12}, in which individuals were allowed to make a choice, before being exposed to the opinions of a group, has revealed that the likelihood of an individual switching their decision depends non-linearly on the proportion of the group who disagree. It is an increasing function which climbs rapidly when the proportion exceeds $50 \%$, also possessing an inflection for large groups. In the context of language, such non-linear conforming behaviour would mean that variants which were used more frequently than others should be used with disproportionately large frequency in the future. A simple way to capture this behaviour is to define
\begin{align}
p_i(\bv{m})&:= \frac{(m_{i})^\beta}{\sum_{j=1}^V (m_{j})^\beta},
\label{eqn:pk}
\end{align}
where $\beta \ge 1$ measures the extent of conformity (non-neutrality). If $\beta=1$ then we have the neutral model, and for $\beta>1$ the non-diffusive term in (\ref{evo}) has a stable fixed point at $m_i=1$.  According to (\ref{eqn:pk}), individuals disproportionately favour the most common variants they have heard: they have a tendency to \textit{conform} to the local majority language use with $\beta$ measuring the strength of this effect. In the limit $\beta \ra \infty$ all speakers use the only the most common (modal) dialect they have heard. An example of this function is plotted in Figure \ref{fig:pFunc} for the $V=2$ model. Conforming behaviour allows \textit{local} dialects to form, as we shall see below. 

The model we have defined is a ``coarse-grained'' description of real linguistic interactions which in reality are much more complex. Much of this complexity arises because there are often many distinct class, ethnic or age-defined social networks in any given geographical region. Within each of these subgroups the need to conform leads to similar speech patterns among members, and these patterns often, but not always, spread to other groups.   Research by linguists has demonstrated that social factors strongly influence the uptake of particular speech patterns \cite{tru00} and that language use is correlated with social class and identity. In American English, for example, language change is often initiated by the working and lower middle classes \cite{lab06,kro78}, before spreading to other groups. Some forms of language change are driven by resistance to conformity; for example ``prestige dialects'' (Received Pronunciation in the UK) are used to signify membership of a social elite, set apart from the common people. The desire to set oneself apart from others can also create reversals in language use amongst subsets of a population. For example, local residents of Martha's Vineyard \cite{lab01,lab63} reverted to an archaic form of pronunciation in order reaffirm local tradition in the face of invading tourists. A similar effect was observed on the island of Ocracoke in North Carolina \cite{wol96}, but in this case the reversal was temporary. As well as social factors, language use may also be determined by age, gender or ethnicity \cite{lab72}. It is clear that reality is far more complex than our simple model, which does not make any of these distinctions between speakers. However, the fact that dialects exists is itself evidence that in general people do adapt to local speech patterns. To model every speaker as having the same need to conform is therefore a reasonable first approximation to reality. It also has the value of simplicity, allowing us later to determine the importance of various additional levels of complexity by comparing how effectively our model fits empirical data when compared more complex models.


\section{Synthetic Dialect Maps}

\label{maps}

\subsection{Application to Great Britain}

\begin{figure}
	\centering
	\includegraphics[width=.8\linewidth]{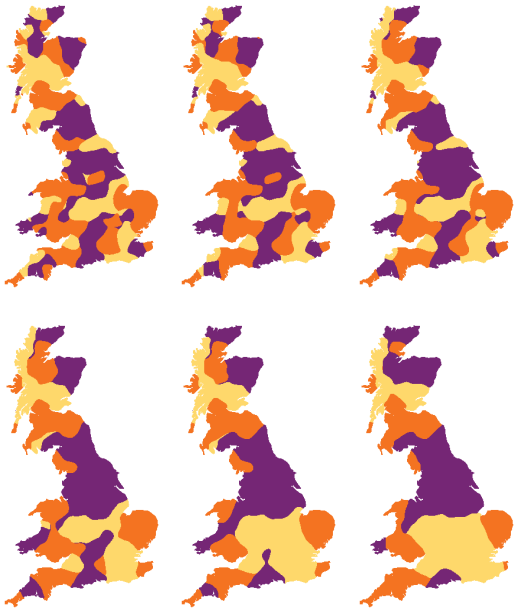}
	\caption{Evolution of the $V=3$ model from randomized initial condition with $\sigma=15$km and $\beta=1.1$ at times $t \in \{1,2,4,8,16,32\}$, where one time unit corresponds to one memory length. Colours indicate which variant is most common at each position. Numerical solution implemented in C++ on grid with 2km spacing \cite{pre07} (GB is $\approx 1000$km north to south). Each grid point initialized with randomly selected variant. }
	\label{fig:evo}
\end{figure}

We apply our model to the island of Great Britain (GB), whose early inhabitants were known as Britons, and spoke Celtic languages \cite{mac08}. The earliest form of English was brought to the island by invading Germanic-speaking settlers. This became \textit{Anglo Saxon} (or \textit{Old English}), as written by Alfred, King of Wessex (849-899 A.D.) but would not be recognisable to modern speakers. It slowly changed, with external influences (notably Norman), into the English we know today \cite{cry03}. 

We seek to discover the extent to which the spatial distribution of dialect structures which have emerged in GB can be predicted by equation (\ref{evo}). To model the evolution of individual linguistic variables we take mainland GB as our spatial domain, and numerically solve equation (\ref{evo}) on a grid of discrete points (Figure \ref{fig:evo}), using an explicit Euler scheme \cite{pre07} (appendix \ref{discrete}). The initial condition for the solution is a randomly generated spatial frequency distribution where each grid point is assigned a randomly selected variant. By repeatedly generating initial conditions and solving the system, we can determine the most probable equilibrium spatial distributions of language use. The population density $\rho(x,y)$ is estimated using 2011 census data \cite{cen11}, which gives the number of inhabitants at each  of the $ \approx 1.8\times 10^6$ UK postcodes. A smooth density is then obtained from this by allowing the inhabitants to diffuse a short distance from the geographical centre of their postcode.  Despite significant overall population growth, the locations of major population centres in GB can trace their origins back through hundreds of years. Since dialect evolution equation (\ref{evo}) depends only on relative population densities, the current density distribution therefore serves as reasonable proxy for historical versions. We estimate that $\sigma$ lies in the range $5\text{km} < \sigma < 15\text{km}$ based on that fact that the average distance travelled to work in GB in 2011 was 15km \cite{cen11}, whereas the average distance travelled to secondary school was 5.5km \cite{nts14}. In section  \ref{analysis} we find that the typical width of a transition region between linguistic variables is  $\approx 1.8 \sigma (\beta-1)^{-\frac{1}{2}}$. For example, the transition between northern and southern GB dialects is $\approx 60km$ wide \cite{cha98}, which if $\sigma = 10km$, gives the approximation $\beta \approx 1.1$.

\subsubsection{Evolution of isoglosses}

\begin{figure}
	\centering
	\includegraphics[width=.8\linewidth]{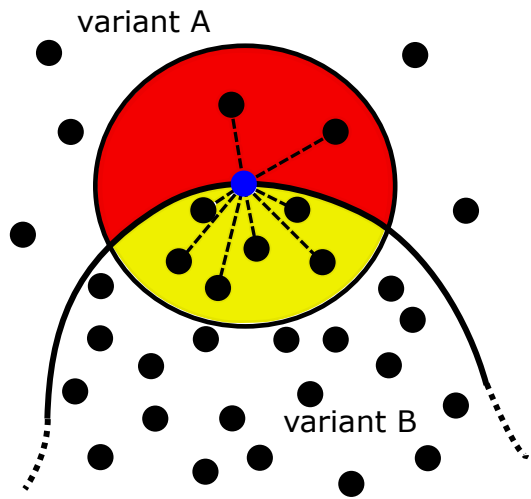}
	\caption{Behaviour of an isogloss surrounding a densely populated area. The blue dot represents a speaker on the isogloss between variants \textbf{A} and \textbf{B}. Other speakers are shown as black dots. The circle around our speaker represents her typical range of interaction. Within red shaded part of this interaction range variant \textbf{A} is more common, and in the yellow shaded part variant \textbf{B} is more common. Due to variation in population density, she hears more of variant \textbf{B} (dashed lines indicate interactions) despite the fact that a greater \textit{area} (red shaded) of her interaction range lies within the domain of variant \textbf{A}.  }
	\label{fig:repulsion}
\end{figure}

When it comes to interpreting our results, the fact that usage frequencies are continuously varying through space presents a similar problem to that faced by dialectologists when trying to draw isoglosses. We resolve this by defining domain boundaries to be lines across which the \textit{modal} (most common) variant changes.  A domain is therefore a region throughout which a single variant is the most commonly used. We may think of domain boundaries as synthetic isoglosses generated by equation (\ref{evo}). 
\begin{figure}
	\centering
	\includegraphics[width=.8\linewidth]{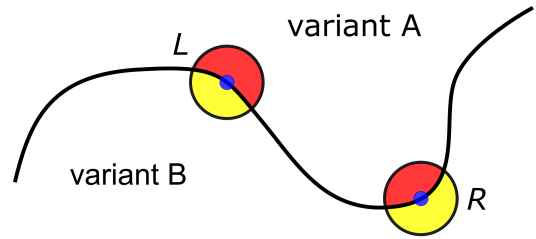}
	\caption{The surface tension effect at domain boundaries. Blue dots represent speakers and black circles give an approximate representation of interaction ranges. In the red shaded parts of these interaction ranges variant \textbf{A} is more common, and in the yellow shaded parts variant \textbf{B} is more common.}
	\label{fig:tension}
\end{figure}
In Figure \ref{fig:evo} we show a series of snapshots of the evolution of domains when there are $V=3$ variants. Isogloss evolution is driven by a two dimensional form of surface tension \cite{deg04}: in the absence of density variation, curved boundaries straighten out. Figure \ref{fig:tension} illustrates why this happens faster when curvature is greater. Here, speaker L hears more of variant A so domain B will retract in this locality. Speaker R  hears more of variant B and so domain A will retract in this region. The net effect will be to straighten the boundary, reducing its length. If a boundary forms a closed curve then this length reduction effect can cause it to evolve toward a circular shape, and reduce in area, eventually disappearing altogether. However, this \textit{shrinking droplet} effect can be arrested or reversed if the droplet surrounds a sufficiently dense population centre (a city). In fact, population centres typically repel isoglosses in our model, and so have a tendency to create their own domains. An explanation of this effect is given in Figure \ref{fig:repulsion}. Here we have a region of high population density in which linguistic variant B is dominant, surrounded by a low population density region where variant A is in common use. We consider the linguistic neighbourhood of a speaker located on the isogloss separating the two domains. From Figure \ref{fig:repulsion} we see that although the majority of the speaker's interaction range lies in region A, she has many more interactions with those in region B, and is therefore likely to adapt her speech toward variant B, causing the isogloss to shift outward into the low density area.

\begin{figure}
	\centering
	\includegraphics[width=.8\linewidth]{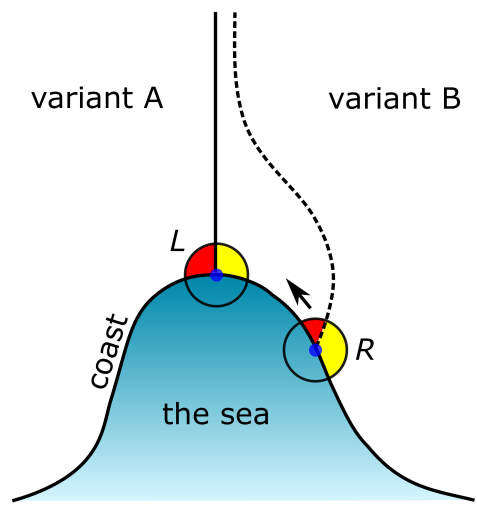}
	\caption{Behaviour of isoglosses at an indentation in coastline or other boundary (political or naturally occurring). Dashed isogloss is unstable and will evolve toward the solid isogloss which emerges perpendicular from the coast. We assume that both isoglosses are effectively anchored to a feature some distance away, opposite the boundary shown. Speakers are shown as blue dots and colours have the same meanings as in Figure \ref{fig:tension}.}
	\label{fig:coast}
\end{figure}

The most common form of stable isogloss generated by our model is a line, typically with some population density induced curvature, connecting two points on the boundary of the system. In order to be stable, such lines must emerge perpendicular from system boundaries, and as a result they are attracted to indentations in coastline, as illustrated in Figure \ref{fig:coast}. In this figure we consider two speakers located at the points where two possible isoglosses meet the coast (or other system boundary - a country border, or a mountain range for example). Speaker R, on the dashed isogloss, hears more of variant B because the isogloss is not perpendicular to the coast; it will therefore migrate upward toward the apex of the coastal indentation until it reaches the stable form shown by the solid line. This effect can be seen in Figure \ref{fig:evo}, where the longest east-west isogloss has migrated so that it emerges from the largest indentation on the east coast of GB. In reality this indentation, called `The Wash', is the site of an \textit{isogloss bundle} (the coincidence of several isoglosses) separating `northern' [\lu] from `southern' [\su] \cite{upt12}. A similar bundle occurs at the largest indentation in the Atlantic coast of France: the Gironde estuary \cite{cha98}, separating the \textit{langue d'oc} from the \textit{langue d'oil}. The fact that bundling at such locations is predicted by our model provides the first sign of the predictive power of the surface tension effect.

\begin{figure}
	\centering
	\includegraphics[width=.9\linewidth]{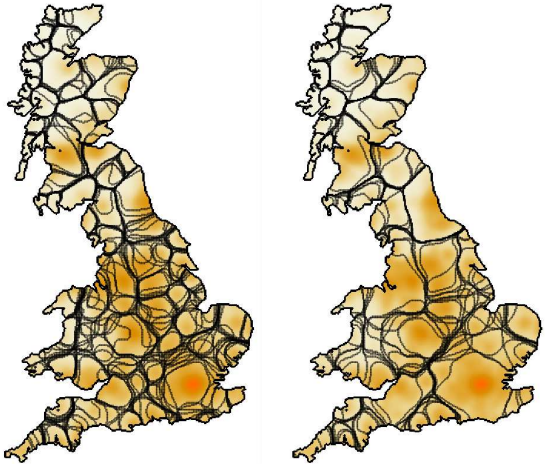}
	\caption{Superposition of the isoglosses at $t=50$ produced by 20 solutions of in the  $V=2$ model with $\beta=1.1$, each with different randomized initial conditions. For the left hand map $\sigma=5$km and for the right hand map $\sigma=10$km (see supplementary video). Background shading indicates population density with brightest orange corresponding to $7200$ inhabitants per km$^2$. }
	\label{fig:iso}
\end{figure}

Having considered the evolution of a single linguistic variable, we now turn to modelling dialects. A dialect is typically defined by multiple linguistic characteristics, and we can capture this by combining many solutions to equation (\ref{evo}). In Figure \ref{fig:iso} we have superposed the synthetic isoglosses for twenty binary ($V=2$) linguistic variables.
We see that there is a significant degree of \textit{bundling} where many isoglosses follow similar routes across the system. Given that the initial conditions for each variable are distinct random frequency distributions (Figure \ref{fig:evo}), then these bundles represent highly probable isogloss positions: many different early spatial distributions lead to these at later stages of evolution. The key point here is that the final spatial structure of dialect domains is rather insensitive to the early history of the language: the effects of surface tension and population density draw many different isoglosses toward the same stable configurations. In this sense the surface tension effect is an `invisible hand' which, in the long term, can overpower historical population upheavals. However, we emphasise that our model predicts only the spatial structure of dialects and not their  particular \textit{sound}; this is very much determined by quirks of history and the initial state of the system. Figure \ref{fig:iso} also illustrates the effect of human mobility on dialect structure. For a smaller interaction range (5km), the structure of synthetic isogloss bundles is more complex, producing a larger number of distinct regions. This effect is well documented in studies of the historical evolution of dialects which were, in the past, more numerous and covered smaller geographical areas \cite{tru99}. Within our model, this is explained by the fact that fluctuations in population density only become relevant to isogloss evolution when they take place over a length scale which is comparable to the interaction range: Two human settlements could only develop distinct dialects if they were separated by a distance significantly greater than $\sigma$, otherwise they would be in regular linguistic contact.  

\begin{figure}
	\centering
	\includegraphics[width=\linewidth]{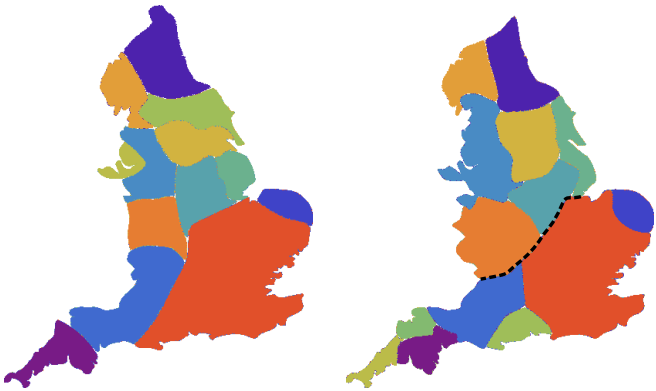}
	\caption{Left map: future England dialect boundaries predicted by Trudgill (1999) \cite{tru99}.  Right map: future dialect boundaries predicted using k-Medoids cluster analysis of 20 synthetic binary linguistic variables when $\sigma=10$km and $\beta=1.1$ at $t=150$. Levenshtein distance (or `edit distance') \cite{lev66} used as distance metric. Colours, determined by Hungarian method, show mapping between dialect areas. Black dotted line shows North-South isogloss.  }
	\label{fig:clust}
\end{figure}	

\begin{figure}
	\centering
	\includegraphics[width=\linewidth]{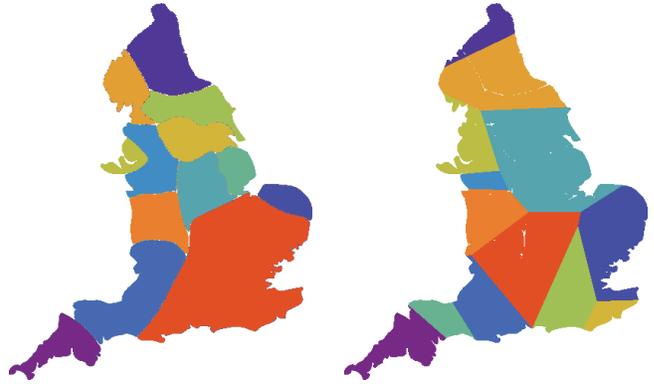}
	\caption{Left map: future England dialect boundaries predicted by Trudgill (1999) \cite{tru99}.  Right map: Voronoi tessellation with the same number of cells as Trudgill's prediction.  Colours determined by Hungarian algorithm.   }
	\label{fig:voron}
\end{figure}	

\subsubsection{Cluster analysis}

Having analysed our model using isoglosses, we now make comparison to recent work in dialectometry, where dialect domains have been determined using cluster analysis and by multi-dimensional scaling \cite{hee04}. A typical clustering approach \cite{sha07,wie11} is to construct a data set giving the frequencies of a wide range of variant pronunciations at different locations, and then to cluster these locations according to the similarity of their aggregated sets of characteristics.  Resampling techniques such as bootstrap \cite{jai88} may be used to generate ``fictitious'' datasets and improve stability.  We mimic this approach by constructing a synthetic data set from 20 solutions of equation (\ref{evo}) with $V=2$, and each with different random initial conditions,  corresponding to different linguistic variables.  We then randomly select a large number (6,000) of sample locations within GB and determine the modal variants for each of the 20 variables at each location. This sample size was chosen to be sufficiently large so that the effect of resampling was only to make short length scale ($\ll$ 1km) changes to cluster boundaries. This aggregated data is then divided into $k$ clusters using the  k-medoids algorithm \cite{mae16} (available in the R language). The metric used for linguistic distance between sample points is the Manhattan distance between the binary vectors where the two variants are labelled 1 or -1. Because we are comparing vectors which can be transformed into one another purely by substitutions (1 for -1 or vice versa), rather than insertions or deletions, then this is equivalent to the Levenshtein distance used in dialectometry \cite{lev66,hee01}. We have found that almost identical results are obtained by applying Ward's hierarchical clustering algorithm \cite{eve11} to the sample locations and subsequently cutting the tree into $k$ clusters.   

\begin{figure}
	\centering
	\includegraphics[width=\linewidth]{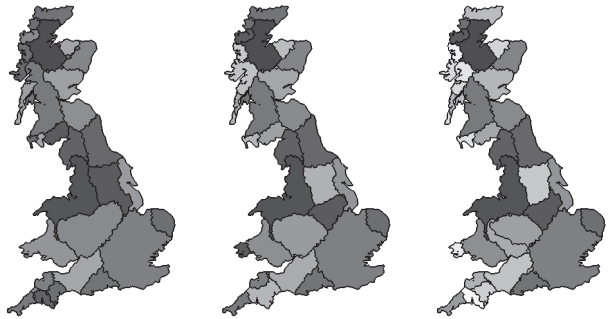}
	\caption{Results of k-Medoids cluster analysis of 20 synthetic binary linguistic variables when $\sigma=10$km and $\beta=1.1$ at $t=150$. Edit distance \cite{lev66} used as distance metric. $k$ values from left to right are $k=16,22,30$.    }
	\label{fig:clusters}
\end{figure}

In order to compare our cluster analysis to the work of dialectologists we consider a prediction for the future dialect areas of England (excluding Wales and Scotland) made by Trudgill \cite{tru99}, shown in the left hand map of Figures \ref{fig:clust} and \ref{fig:voron}. This prediction divides the country into 13 regions, and is the result of a systematic analysis of regional variation in speech and ongoing changes. Such sharp divisions are a significant simplification of reality however, and hide many subtle smaller scale variations. The decision to define 13 regions therefore reflects a judgement on the range of language use which can be categorized as a single dialect. To allow comparison with this prediction we have performed a set of cluster analyses of near-equilibrium (large $t$) solutions for the whole of GB, for a range of values of the number, $k$, of clusters (see Figure \ref{fig:clusters}), with the aim of producing 13 within the subset of GB defined by England. The closest result was 14 clusters for $20\leq k \leq 24$ with almost identical results within England for each of these choices. Having defined our synthetic dialect regions we apply the Hungarian method \cite{kuh55} to find the mapping between our synthetic dialects, and Trudgill's predicted dialects, which maximizes the total area of overlap between the two. The results are show in Figure \ref{fig:clust}. To provide a measure of the effectiveness of our model in matching Trudgill's predictions we also define a \textit{null} model, which divides the country into regions at random, independent of population distribution and without reference to any model of speaker interaction. There are a number of models which generate random tessellations of space \cite{chi13}, many of which are motivated by physical processes such as fracture or crack propagation. We wish to exclude such physical assumption and so opt for the Voronoi tessellation \cite{chi13}, based on the Poisson point process: the simplest of all random spatial processes. Our null model is then a Voronoi tessellation of England (Figure \ref{fig:voron}) using 13 points selected uniformly at random from within its borders, with dialects labelled to most closely match Trudgill's map, using the Hungarian method.

\begin{table}
	\caption{\label{tab:sim} Metrics measuring the similarity between Trudgill's predictions \cite{tru99} for future English dialects and the predictions of our model. Metric acronyms are OL (Overlap), WOL (Weighted Overlap), RI (Rand Index) and ARI (Adjusted Rand Index). The \textbf{Voronoi Example} column gives equivalent metrics for the example Voronoi tessellation in Figure \ref{fig:voron} and the \textbf{Voronoi Set} column gives the mean metrics, with standard deviation, for five randomly generated Vornoi tessellations. }
	\begin{ruledtabular}
		
		\begin{tabular}{|c|c|c|c|}
			\hline 
			\textbf{Metric} & \textbf{Model}  & \textbf{Voronoi Example} & \textbf{Voronoi Set} \\ 
			\hline 
			\text{OL} & 68\%  & 41\% & $45 \pm 3\% $\\ 
			\hline 
			\text{WOL} & 82\%  & 36\%  & $49 \pm 11\%$ \\ 
			\hline 
			\text{RI} & 0.91 & 0.84 & $0.83 \pm 0.01$ \\ 
			\hline 
			\text{ARI}& 0.63 & 0.29 & $0.30 \pm 0.01$   \\ 
			\hline 
		\end{tabular} 
	\end{ruledtabular}
\end{table}

Having generated a our synthetic dialect maps, we now quantify the extent to which they match the predictions of Trudgill. The null model, because its lack of modelling assumptions, will reveal the extent to which our model is ``better than random'' at matching these predictions.  We offer four alternative metrics of similarity in Table \ref{tab:sim}. The simplest metric is overlap (OL): the percentage of land area which is identified as belonging to the same dialect as Trudgill's prediction. The \textit{Weighted Overlap} (WOL) weights overlapping regions in proportion to their population density: it gives the probability that a randomly selected inhabitant will be assigned to the same dialect zone by both maps. From table \ref{tab:sim} we see that this probability is high $(82 \%)$ for our model, but lower for the random Voronoi model. We suggest that this is a result of the fact that population centres tend to \textit{repel} isoglosses and therefore often lie at the centres of dialect domains. We will examine this repulsion effect in more detail below. The final two metrics are commonly used to compare clusterings. Consider a set, $S$ of elements (spatial locations for us) that has been partitioned into clusters (dialect areas) in two different ways. Let us call these two partitions $X$ and $Y$. The Rand Index (RI) \cite{ran71} is defined as the probability that given two randomly selected elements of $S$, the partitions $X$ and $Y$ will agree in their answer to the question: are both elements in the same cluster? A disadvantage with using this index to compare dialect maps is that the larger the number of regions in the maps, the more likely it is that two randomly selected spatial points will not lie in the same cluster in either map. The index therefore approaches 1 as the number of dialect areas grows. This problem may be countered by taking account of its expected value if $X$ and $Y$ were picked at random, subject to having the same number of clusters and cluster sizes as the originals \cite{hub85}. The ``Adjusted Rand Index'' (ARI) is then defined 
\begin{equation}
\text{ARI} := \frac{\text{RI}-\text{expected index}}{1-\text{expected index}}.
\end{equation}
The ARI $\in [-1,1]$ measures the extent to which a clustering is a better match than random to some reference clustering and is used by dialectometrists \cite{her00} in preference to the original Rand Index (RI). For us the reference clustering is Trudgill's predicted dialect map, and the Rand and Adjusted Rand Indices in Table \ref{tab:sim} measure similarity to this reference.  

The primary conclusion that may be drawn from the indices in Table \ref{tab:sim} is that by all measures our model provides a better match than the null model (indices all differ by at least three standard deviations, and typically many more). Of particular interest is the weighted overlap probability (WOL= 82\%). 
Isoglosses are typically repelled by population centres, so tend to pass through regions of relatively low density. Because of this the WOL may be viewed as a measure of the effectiveness of the model at determining the centres of dialect regions and is less sensitive to small errors in isogloss construction, explaining its high value. It is important to realise also that Trudgill's predictions may themselves be imperfect.

We now make some qualitative comments. 
The dotted line in Figure \ref{fig:clust} shows the location of our model's most dense North-South isogloss bundle. This is coincident with what is described by Trudgill as ``one of the most important isoglosses in England'' \cite{tru74} dividing those who have [\lu] in butter from those who do not. In our model, the fact that this border lies where it does is a result of the surface tension effect which attracts many isoglosses towards the two coastal indentations at either end (see supplementary video). 
The fact that many randomized initial boundary shapes evolve toward this configuration, and that the configuration is seen as important by dialectologists supports the hypothesis that surface tension is an important driver of spatial language evolution. We also note that the western extremities of GB (Cornwall and North-West Scotland) support multiple synthetic dialects in our model, and we suggest that this is due to a heavily indented coastline and the fact that high aspect-ratio tongues of land are likely to be crossed by isoglosses; a fact predictable by analogy with continuum percolation \cite{bar09}. The south west peninsula has historically supported three dialects. 

In future work, the model might be tested by comparing its predictions to well researched dialect areas. On example is the Netherlands. Here, dialectologists have performed a cluster analysis \cite{hee04} based on Levenshtein distances between field observations of 360 dialect varieties (correspoding to 357 geographical locations), revealing 13 significant geographical groupings. The extent to which a model is consistent with these groupings, accounting for variability caused by finite sample size, could be tested by generating equivalent clusterings for multiple fictitious dialect samples of the same size.    


\subsection{Bundles, Fans, Stripes and Circular Waves}

\begin{figure}
	\centering
	\includegraphics[width=\linewidth]{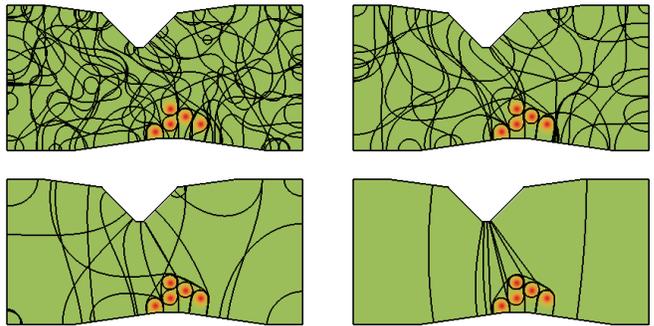}
	\caption{ Evolution of isoglosses in a $400 \times 200$ system with two opposing boundary indentations and unit background population density, together with a collection of cities contributing addition population densities $\rho(r) = (\rho_1-1) \exp\left\{-r^2/(2 R^2)\right\}$ where $r$ is distance from city centre, $R=10$ and $\rho_1=4$ (measuring ratio of peak city density to background). Parameter values: $\sigma = 4$, $\beta = 2$. Evolution times $t = 10, 50, 300, 1660$. See supplementary video for full animation.  }
	\label{fig:fan}
\end{figure}	

We now illustrate a number of well known features of dialect distributions which may be qualitatively reproduced by our model.  
We consider first the isogloss bundle reported by Bloomfield (1933) \cite{blo33} separating ``High German'' from ``Low German''. The bundle emerged from the tip of an indentation of the Dutch-German speech area (bordered to the East by Slavic languages), and ran roughly East-West before separating approximately 40km East of the river Rhine, and fanning out around cities such as Dusseldorf, Cologne, Koblenz and Trier. This arrangement of isoglosses is known as the ``Rhenish Fan''. In Figure \ref{fig:fan} we have constructed an artificial system with boundaries approximating the geographical structure of relevant parts of the Dutch-German language area illustrated in Bloomfield \cite{blo33} containing an artificial cluster of population centres representing the German cities located near to the Rhine. The system was initialized using the same randomization procedure used for GB, and Figure \ref{fig:fan} shows a superposition of ten solutions, each with different initial conditions. In the early stages of evolution, very little pattern is discernible, but as time progresses the main indentation collects isoglosses, while the cities repel them, producing a fan-like structure. We therefore suggest that the isogloss separation which created the Rhenish fan may have been the result of repulsion by the cities of the  Rhine.

\begin{figure}
	\centering
	\includegraphics[width=\linewidth]{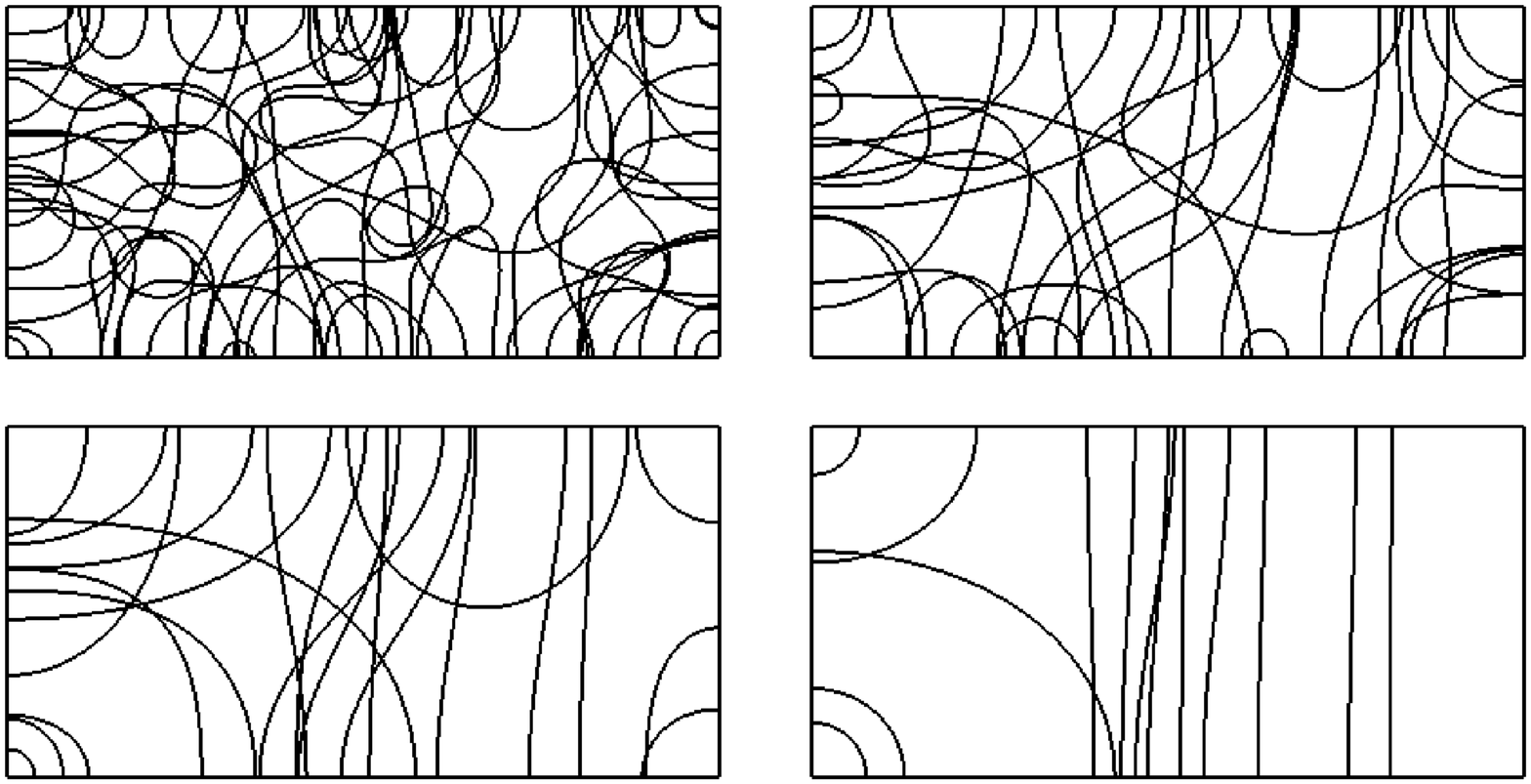}
	\caption{ Evolution of isoglosses in a $400 \times 200$ rectangular system with uniform population density, starting from randomized initial conditions. Figure shows a superposition of 10 such solutions. Notice that all isoglosses join the two long sides of the system. Parameter values: $\sigma = 4$, $\beta = 2$. Evolution times $t = 30, 90, 270, 810$.   }
	\label{fig:stripe}
\end{figure}	

We next consider an example of what some physicists refer to as ``stripe states'' \cite{bar09}: in finite systems which experience phase ordering, and have aspect ratios greater than one, boundaries between two orderings often form across the system by the shortest route (in a rectangle, joining two long sides).  A collection of such boundaries form a striped pattern of phase orderings. Figure \ref{fig:stripe} illustrates this effect, produced by equation (\ref{evo}). Our model therefore predicts such striped dialect patterns in long thin countries, and a particularly striking example of the effect may be seen in the dialects of the Saami language \cite{sam98}.  The Saami people are indigenous to the S\'{a}mpi region (Lapland) which includes parts of Norway, Sweden and Finland. Their Arctic homeland forms a curved strip with a length which is approximately five times its average width. The region is divided into ten language areas, and the boundaries of all but two of these take a near-direct route between the two long boundaries of the system, forming a distinctive striped pattern.   Another example are the dialects of Japan whose boundaries in many cases cross the country perpendicular to its spine \cite{pre99}.

\begin{figure}
	\centering
	\includegraphics[width=\linewidth]{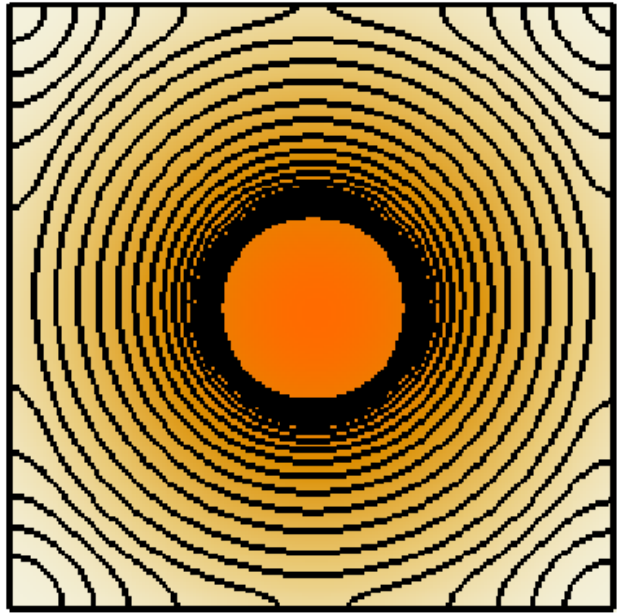}
	\caption{ Evolution of a circular isogloss (initial radius 30) in a $200 \times 200$ system with unit background population density, together with a central city contributing additional density $\rho(r) = (\rho_1-1) \exp\left\{-r^2/(2 R^2)\right\}$ where $r$ is distance from city centre, $R=40$ and $\rho_1=21$. Parameter values: $\sigma = 4$, $\beta = 2$. Evolution times $t \in \{10, 20, 30 \ldots, 300\}$.   }
	\label{fig:city}
\end{figure}

The relationship between geographical separation and linguistic distance (often measured using Levenshtein distances \cite{hee01}) is typically sublinear \cite{hee01,ner10}. The definition of linguistic distance and its relation to geographic distance was made by S\'{e}guy \cite{seg71,seg73}, and the relationship therefore goes by the name: ``S\'{e}guy's Curve'' \cite{ner10}. It has been substantially refined and tested since \cite{hee01,ner10,spr09}, and also generalised to involve travel time \cite{goo05}. S\'{e}guy's curve is not universal, however. For example, an analysis of Tuscan dialect data \cite{mon08} reveals an unusually low correlation between phonetic and geographical distances. A more detailed analysis reveals that there are geographically remote areas which are linguistically similar, and that within an approximately circular region (radius $\approx 40\text{km}$) around the main city, Florence,  phonetic variation correlates more strongly with geographical proximity. It is hypothesized \cite{mon08} that this pattern marks the radial spread of a linguistic innovation (called ``Tuscan -gorgia''). This Tuscan data motivates our final example of the qualitative behaviour of our model. To illustrate how a linguistic variable can spread outwards from a population centre, purely through the effects of population distribution and not necessarily driven by prestige or other forms of bias, we have simulated our model using an artificial city with Gaussian population distribution (Figure \ref{fig:city}). The system is initialized with a circular isogloss, centred on the city, representing a local linguistic innovation. Because population density is a decreasing function of the distance from the city centre, speakers on the isogloss hear more of the innovation than their current speech form, allowing it to expand (as explained in Figure \ref{fig:repulsion}). We will see in section \ref{analysis} that this expansion will not not necessarily continue indefinitely. Expansion processes such as this have also been observed in Norway \cite{tru74}. In that case the progress of new linguistic forms was shown to depend on age, with changes more advanced for younger speakers, who are more susceptible to new form of speech. We illustrate how this effect can be analysed in  appendix \ref{age}.

\section{S\'{e}guy's Curve}
\label{seg}

We now determine the relationship between geographical and linguistic distance within our model, providing an analytical prediction for the form of S\'{e}guy's curve \cite{ner10,seg71,seg73}. For simplicity we consider the two variant model $V=2$ and suppose that our language contains a number, $n$, of linguistic variables. At each location in space the local dialect is an $n$ dimensional vector of the local modal variants which we label $1$, and $-1$. Letting $\bv{\phi}(\bv{r},t)$, where $\bv{r}=(x,y)$, be the vector field giving the distribution of these variants then the number of differences (the Levenshtein distance \cite{lev66}) between two dialects $\bv{\phi}(\bv{r}_1,t)=: \bv{\phi}(1)$ and $\bv{\phi}(\bv{r}_2,t)=:\bv{\phi}(2)$  is $(n-\bv{\phi}(1) \cdot \bv{\phi}(2))/2$. The linguistic distance, $L(1,2)$, between two dialects may be defined \cite{ner03} as the fraction of variables that differ between them
\begin{equation}
L(1,2) := \frac{1}{2}\left(1-\frac{\bv{\phi}(1) \cdot \bv{\phi}(2)}{n} \right).
\end{equation}
Since we have assumed that each variant evolves independently of every other, then the expected linguistic distance is
\begin{equation}
l(1,2) := \EE[L(1,2)] = \frac{1}{2}\left(1-\EE[\phi_i(1) \phi_i(2)] \right),
\end{equation}
where $\phi_i$ is the $i$th component of $\bv{\phi}$. To compute $\EE[\phi_i(1) \phi_i(2)]$ we make use of the close similarity between equation (\ref{evo}) and the time dependent Ginzburg Landau equation \cite{bra94}, to derive (see appendix \ref{AC}) an analogue of the Allen-Cahn equation \cite{all79} giving the velocity of an isogloss at a point in terms the unit vector $\unit{g}$ normal to it at that point
\begin{align}
v &= -\beta \sigma^2 \left[ \frac{\nabla \cdot \unit{g}}{2} + \frac{\nabla \rho \cdot \unit{g}}{\rho} \right] \\
&= -\beta \sigma^2 \left[ \frac{\kappa}{2} + \frac{\nabla \rho \cdot \unit{g}}{\rho} \right].
\label{eqn:AC}
\end{align}
The quantity $\kappa$ is the curvature of the isogloss at the point: in the absence of variations of population density, the isogloss moves so as to reduce curvature. The second term in the square brackets produces a net migration of isoglosses towards regions of lower population density. To compute correlation functions between the field $\phi_i$ at different locations in space we apply the Ohta-Jasnow-Kawasaki (OJK) method \cite{oht82}, introducing smoothly varying auxiliary field $m(x,y,t)$ which gives the value of the $i$th variant as $\phi_i = \text{sgn}(m)$. Note that the auxiliary field $m(x,y,t)$ is distinct from the memory $m_i(x,y,t)$ for the $i$th variant.
The OJK equation, describing the time evolution of this field, adapted to include density effects, is (appendix \ref{AC})
\begin{equation}
\frac{\p m}{\p t} =  \beta \sigma^2 \left(\frac{\nabla^2 m}{4} + \frac{\nabla \rho.\nabla m}{\rho} \right).
\label{OJK}
\end{equation}
We introduce the fundamental solution, $G(\bv{r},t;\bv{r}_0)$, (the Green's Function)  of (\ref{OJK}), giving the function $m(\bv{r},t)$ subject to the initial condition $m(\bv{r},0)=\delta(\bv{r}-\bv{r}_0)$. The solution for arbitrary boundary conditions is then
\begin{equation}
m(\bv{r},t) = \int_{\mathbb{R}^2} d\bv{r}_0 G(\bv{r},t;\bv{r}_0) m(\bv{r}_0,0).
\end{equation}
We assume that the initial condition of our system consists of spatially uncorrelated language use, so that $\EE[\phi_i(\bv{r}_1,0) \phi_i(\bv{r}_2,0)] = \delta_{\bv{r}_1 \bv{r}_2}$. A convenient, equivalent condition on the auxiliary field is to let it be Gaussian (Normally) distributed $m(\bv{r},0) \sim \mathcal{N}(0,1)$ with correlator   
\begin{equation}
\EE[m(\bv{r}_1,0) m(\bv{r}_2,0)] = \delta(\bv{r}_1-\bv{r}_2).
\end{equation}
We can compute this correlator at later times using the fundamental solution $G$
\begin{multline}
\EE[m(\bv{r}_1,t) m(\bv{r}_2,t)]  \\ 
= \int_{\mathbb{R}^4} G(\bv{r}_1,t;\bv{r_1}') G(\bv{r}_2,t;\bv{r_2}')\EE \left[ m(\bv{r}_1',0) m(\bv{r}_2',0) \right] d \bv{r}_1' d \bv{r}_2' \\
= \int_{\mathbb{R}^2} G(\bv{r}_1,t;\bv{r}_0) G(\bv{r}_2,t;\bv{r}_0) d \bv{r}_0. 
\end{multline}
The linearity of our adapted OJK equation (\ref{OJK}) ensures that $m(\bv{r},t)$ remains Gaussian for all time \cite{bra94} (to see this note that derivatives are limits of sums, and sums of Gaussian random variables are themselves Gaussian). However, the values of the field at different spatial locations develop correlations so that the joint distribution of any pair is bivariate normal. Following Bray \cite{bra94} we define the normalized correlator
\begin{equation}
\gamma(\bv{r}_1,\bv{r}_2) := \frac{\EE[m(\bv{r}_1,t) m(\bv{r}_2,t)]}{\sqrt{\EE[m(\bv{r}_1,t)^2]. \EE[m(\bv{r}_2,t)^2]}}.
\end{equation}
Using the abbreviated notation $\gamma(\bv{r}_1,\bv{r}_2) \equiv \gamma(1,2)$, the correlator for the original field may be found by averaging over the bivariate normal distribution of $m(\bv{r}_1,t)=:m(1)$ and $m(\bv{r}_2,t)=:m(2)$ (see, e.g. \cite{oon88})
\begin{align}
\EE[\phi_i(1) \phi_i(2)](t) &= \EE[\sgn [m(1)] \sgn[m(2)]] \\
&= \frac{2}{\pi} \sin^{-1}(\gamma(1,2)).
\label{eqn:correl}
\end{align}
We now compute this correlator and derive a theoretical prediction for S\'{e}guy's Curve.

\subsection{Uniform population density}

\begin{figure}
	\centering
	\includegraphics[width=\linewidth]{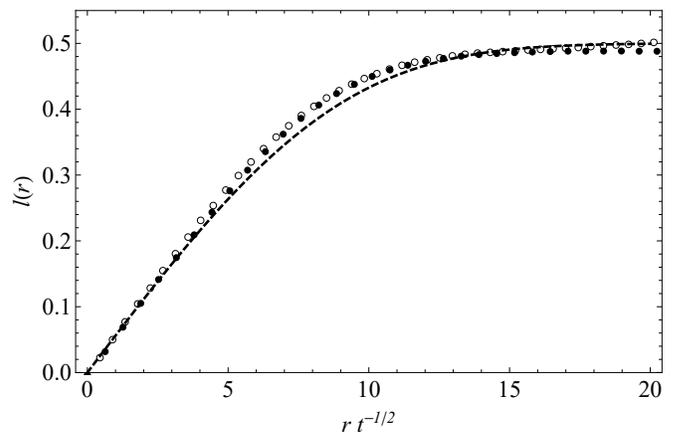}
	\caption{ S\'{e}guy's curve showing linguistic distance ($l$) versus geographical distance ($r$). Dashed line shows equation (\ref{eqn:seg}) in the case $\sigma=4, \beta=2$. Open and closed dots show simulated linguistic distances using same parameter values in a $1000\times1000$ system at times $t=80,160$. Note that linguistic distance depends only on the combination $rt^{-1/2}$ so curves evaluated at different time collapse onto one another.   }
	\label{fig:seg}
\end{figure}	

If population density is constant $\rho=C$ then our adapted OJK equation (\ref{OJK}) reduces to OJK's original form which has fundamental solution
\begin{equation}
G(\bv{r},t;\bv{r}_0) = \frac{\exp\left\{- \frac{|\bv{r}-\bv{r}_0|^2}{\beta \sigma^2 t}\right\}}{\pi \beta \sigma^2  t}
\end{equation}
giving a normalized correlator
\begin{equation}
\gamma(1,2) = \exp\left\{-\frac{r^2}{2t \beta \sigma^2 } \right\} =:\gamma_t(r).
\end{equation}
Our prediction for S\'{e}guy's curve at time $t$ is therefore
\begin{equation}
l(r,t) = \frac{1}{2} \left[1-\frac{2}{\pi} \sin^{-1}(\gamma_t(r))\right].
\label{eqn:seg}
\end{equation}
This curve is plotted in Figure \ref{fig:seg} along with simulation results. We give the following interpretation of the curve. Starting from a randomized spatial distribution of language use, the need for conformity generates localized regions where particular linguistic variables are in common use, and these regions are bounded by isoglosses. These regions expand, driven by surface tension in isoglosses, so that from any given geographical point one would need to travel further in order to find a change in language use. The linguistic distance between two points therefore tends to decrease with time, and the curve (\ref{eqn:seg}) gives the rate of decrease as exponential. There are  features of reality which we might expect to alter this behaviour. First, we have assumed that no major population mixing or migration takes place - such events would have the effect of resetting the initial conditions of the model. Our prediction is only valid during times of stability. Second we have assumed that the population is uniformly distributed in the system when in reality populations are clumped and, as we have seen, population centres can support their own dialects if they are large enough. We take some steps toward addressing this issue below. In appendix \ref{ner} we briefly discuss a simple one dimensional simulation model from the dialectology literature \cite{ner10}, which includes the same large $r$ behaviour as in (\ref{eqn:seg}) for a particular choice (quadratic) of macroscopic ``influence'' curve.

\subsection{Peaked population density}

We now consider how S\'{e}guy's curve is modified by the presence of a peak in population density. In order to allow analytical tractability we consider a simple exponentially decaying peak
\begin{equation}
\rho(x,y) = \exp \left\{ -\frac{\sqrt{x^2+y^2}}{R} \right\}
\label{eqn:expPop}
\end{equation}
where $R>1$. To understand the behaviour of the modified OJK equation (\ref{OJK}) it is useful to decompose it into an advection diffusion equation  plus a source term
\begin{equation}
\frac{\p m}{\p t} =  \beta \sigma^2 \left[ \nabla\cdot \left(\frac{\nabla m}{4} + \frac{\nabla \rho}{\rho} m \right)  - \left(\nabla \cdot  \frac{\nabla \rho}{\rho}\right) m\right].
\label{OJK_AD}
\end{equation}
Defining $r:=\sqrt{x^2+y^2}$, the average velocity field for the diffusing particle is 
\begin{equation}
-\frac{\nabla \rho}{\rho} = \frac{(x,y)}{r R}.
\end{equation}
The source term is 
\begin{equation}
- \left(\nabla \cdot  \frac{\nabla \rho}{\rho}\right) m = \frac{m}{r R}.
\end{equation}
We now view equation (\ref{OJK_AD}) as describing the mass distribution for a collection of Brownian particles which are being driven radially away from the origin. The source term is interpreted as a field which causes particles to produce offspring at rate $(rR)^{-1}$ as they move through it. The fundamental solution, $G(\bv{r},t;\bv{r}_0)$, to (\ref{OJK_AD})  is then the mass distribution for a very large (approaching infinite) collection of particles with total mass initially equal to one, all of which started at $\bv{r}_0$. 

\begin{figure}
	\centering
	\includegraphics[width=\linewidth]{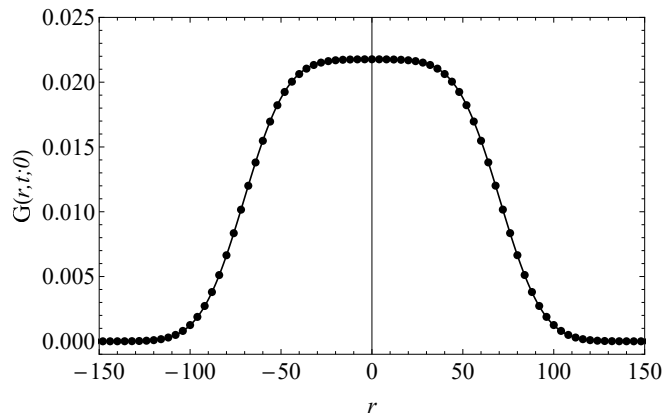}
	\caption{ Continuous line line shows radial cross section (along the line $y=0$) through numerical solution of the modified OJK equation (\ref{OJK_AD}) with population distribution (\ref{eqn:expPop}) at time $t=700$ with initial condition $\bv{r}_0 = (0,0)$. Here $r=\sqrt{x^2+y^2} = x$. Parameter values are $\beta \sigma^2=1$ and $R=10$.  Data points show asymptotic analytical solution $G(|r|,t;0)$ (equation (\ref{eqn:Gana})) with $t=700$, time offset $t_0=-4.4$ and $A=0.0109$ (found by maximum likelihood). }
	\label{fig:OJKAsymp}
\end{figure}

\begin{figure}
	\centering
	\includegraphics[width=\linewidth]{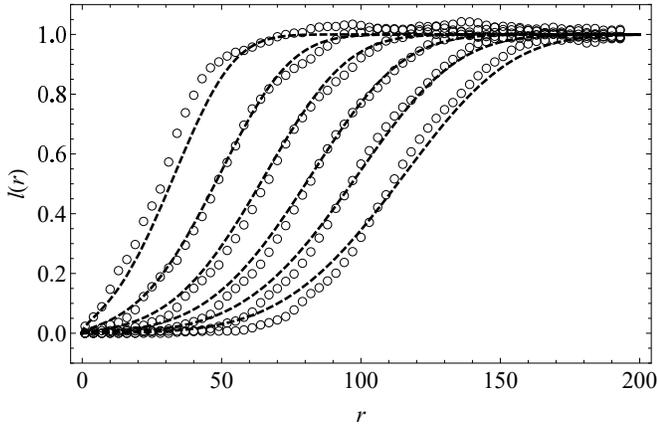}
	\caption{ Dashed lines show theoretical shape of S\'{e}guy's curve (\ref{eqn:segCity}) centred at peak of population density $\rho=e^{-r/R}$ with $R=20$. Curve computed using equation (\ref{eqn:segCity}) when $\beta=1.4, \sigma=5$, times are $t=10,20,30$ with offset $t_0=-5.77$ (maximum likelihood estimate). Simulation points give equivalent correlations in the full model computed from 100 independent simulations in a $400 \times 400$ system.  }
	\label{fig:segCity}
\end{figure}

\begin{figure}
	\centering
	\includegraphics[width=\linewidth]{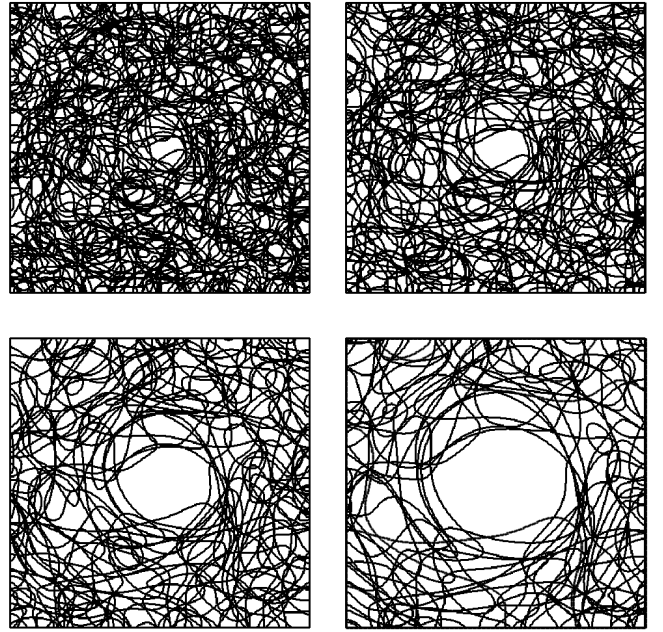}
	\caption{ Isogloss evolution in a  $400 \times 400$ system with $V=2,\beta=1.1,\sigma=5$  at $t=10,15,25,35$ with $\rho=e^{-r/R}$ where $R=20$ and $r=0$ corresponds to the centre of the system. Plot is a superposition of 20 simulations with different initial conditions. Central peak repels isoglosses.  See supplementary video for full animation.  }
	\label{fig:tangle}
\end{figure}

We wish to compute the dependence of linguistic distance on geographical distance from the peak of the population density (thought of as the centre of a city). We therefore require the expectation
\begin{equation}
\EE[m(\bv{0},t) m(\bv{r},t)]  = \int_{\mathbb{R}^2} G(\bv{0},t;\bv{r}_0) G(\bv{r},t;\bv{r}_0) d \bv{r}_0. 
\label{eqn:momr}
\end{equation}
Computation of a general closed form expression for $G(\bv{r},t;\bv{r}_0)$ is not our aim; preliminary computations in this direction suggest that if such a form existed its complexity would restrict its use to numerical computations alone. Instead we make arguments leading to a simple approximation for S\'{e}guy's curve. We observe first that the integrand in (\ref{eqn:momr}) is dominated by the region around $\bv{r}_0=\bv{0}$. 
\begin{figure}
	\centering
	\includegraphics[width=\linewidth]{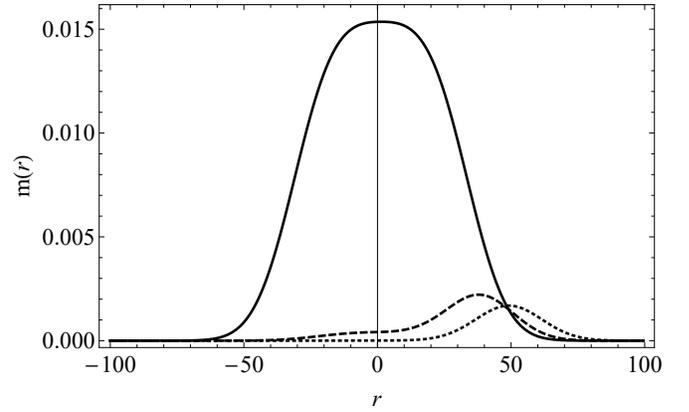}
	\caption{ Radial cross sections (along the line $y=0$) through numerical approximations to fundamental solutions of the modified OJK equation (\ref{OJK_AD}) with population distribution (\ref{eqn:expPop}) at time $t=300$. Here $r=\sqrt{x^2+y^2} = x$. Parameter values are $\beta \sigma^2=1$ and $R=10$. Initial conditions are $\bv{r}_0 = (1,0),(10,0),(20,0)$ (solid, dashed, dotted lines respectively).}
	\label{fig:OJK}
\end{figure}	
Numerical evidence for this is provided in Figure \ref{fig:OJK} where we see that the fundamental solution grows in magnitude as $\bv{r}_0 \ra \bv{0}$. 
In general the solution consists of a circular plateaux propagating outward from the origin plus an isolated but spreading peak also drifting away from the origin (Figure \ref{fig:OJK}). The plateaux is formed once the rate of loss of particles from the peak source region ($|\bv{r}_0| \lesssim R^{-1}$) though advection and diffusion is equal to the rate of the creation of new particles. The plateaux height is determined by the particle mass which reaches the peak source region in the early stages of evolution.
Due to radial drift, the only particles with a chance of doing this are those with sufficiently small P\'{e}clet number \cite{pat80}
\begin{equation}
\mbox{\textit{Pe}} = \frac{4|\bv{r}_0|}{R}, 
\end{equation}
where $\bv{r}_0$ is their starting point (or that of their earliest ancestor if they are daughters). Values of $\bv{r}_0$ which lie outside a region  of radius $ \propto R$  (henceforth \textit{P\'{e}clet region}) around the origin can therefore be ignored when computing $G(\bv{0},t;\bv{r}_0)$.  For $R\gg 1$, the peak region forms a small fraction $O(R^{-4})$ of the P\'{e}clet region and particles within the P\'{e}clet region have a probability of reaching the peak which decays exponentially with their initial distance from it. The function $G(\bv{0},t;\bv{r}_0)$ will therefore itself be sharply peaked within the P\'{e}clet region, around $\bv{r}_0=\bv{0}$, and we make the approximation $G(\bv{0},t;\bv{r}_0) \approx h \delta(\bv{r}_0) $, where $h$ is plateaux height. Making use of this approximation in (\ref{eqn:momr}) we have
\begin{equation}
\EE[m(\bv{0},t) m(\bv{r},t)]  \approx h G(\bv{r},t;\bv{0}).
\end{equation}
To compute the variance
\begin{equation}
\EE[m^2(\bv{r},t)] = \int_{\mathbb{R}^2} G^2(\bv{r},t;\bv{r}_0) d\bv{r}_0
\end{equation}
we note that if $|\bv{r}|\ll t/R$ then the dominant contribution to the integral comes from the plateaux component of the solution. If $|\bv{r}|\gg t/R$ then the plateaux will not have reached $\bv{r}$ so only the spreading peak component of the fundamental solution will contribute. Therefore
\begin{equation}
\sqrt{\EE[m^2(\bv{r},t)]}\approx
	\begin{cases}
	O(1) & \text{ if } |\bv{r}|\ll t/R \\
	O(t^{-1/2}) & \text{ if }  |\bv{r}| \gg t/R.
	\end{cases}
	\label{eqn:var}
\end{equation}
We will comment on the significance of this behaviour below. To find the form of $G(\bv{r},t,0)$ we note first its circular symmetry, which reduces the number of variables in the OJK equation to two
\begin{equation}
\frac{\p m}{\p t} = \frac{\sigma^2 \beta}{4} \left[ \frac{\p^2 m}{\p r^2} + \left( \frac{1}{r} -\frac{4}{R}\right)\frac{\p m}{\p r}  \right].
\label{eqn:circOJK}
\end{equation}
We seek a travelling wave solution, subject to the initial condition $m(r,0)=\delta(r)$, representing the expanding plateaux, valid for large $r$ so that the $1/r$ term in (\ref{eqn:circOJK}) can be neglected. We obtain, as $t \ra \infty$
\begin{equation}
G(r,t;0) \sim A \times \text{erfc}\left(\frac{R  r -\beta  \sigma ^2 (t-t_0)}{R \sigma  \sqrt{\beta (t-t_0)}}\right)
\label{eqn:Gana}
\end{equation}
where $t_0$ is a time correction which accounts for the fact that the propagation velocity of the plateaux takes some time to settle down to its long time value of $\beta \sigma^2/R$. We verify in Figure \ref{fig:OJKAsymp} that this is the correct asymptotic solution by comparing it to the numerical solution of (\ref{eqn:circOJK}) for large $t$.
We now approximate the normalized correlator as 
\begin{equation}
\gamma(r,t) \approx \frac{G(r,t;0)}{G(0,t;0)}.
\end{equation}
This approximation neglects the drop in the variance of $m(r,t)$ for $r \gg t/R$ described by equation (\ref{eqn:var}), which amounts to neglecting a multiplicative factor $\sqrt{t}$ in the large $r$ behaviour of the correlator.  Our approximate analytical prediction for S\'{e}guy's curve measured radially from the centre of the exponentially decaying population distribution is therefore
\begin{equation}
l_c(r,t) := \frac{1}{2} \left[1-\frac{2}{\pi} \sin^{-1}\left(\frac{G(r,t;0)}{G(0,t;0)}\right)\right].
\label{eqn:segCity}
\end{equation}
This prediction is compared to correlations in the full model (Figure \ref{fig:segCity}) by generating 100 realizations of isogloss evolution over the exponential population density, each with different randomized initial conditions. From Figure \ref{fig:segCity} we see that as time progresses a growing region emerges around the centre of the city in which the linguistic distance to the centre is close to zero.
An alternative visualization of this effect is given in Figure \ref{fig:tangle} which shows a superposition of the isoglosses from 20 simulation runs. As time progresses a circular patch emerges in the centre of the system which is devoid of isoglosses, and therefore where all speakers use the same linguistic variables. Outside of this central ``city dialect'' we note that the asymptotic behaviour of the complementary error function
\begin{equation}
\text{erfc}(x) \sim \frac{\exp\{-x^2\}}{x} \text{ as } x \ra \infty,
\end{equation}
together with the expansion $\sin^{-1}(\epsilon) = \epsilon + O(\epsilon^3)$ lead to the prediction that linguistic correlations fall as $e^{- c (\Delta r)^2}/(\Delta r)$ where $c$ is a constant and $\Delta r$ is distance from the edge of the city dialect. This is a faster rate of decay than in the flat population density case. It appears from Figure \ref{fig:seg} that in reality the decay rate may be even faster than this. 
\begin{figure}
	\centering
	\includegraphics[width=\linewidth]{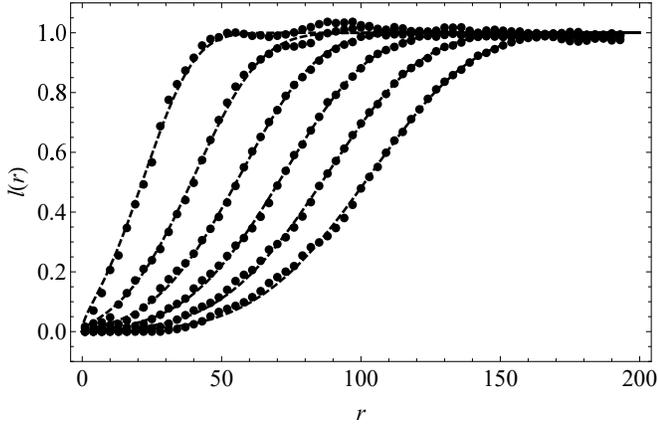}
	\caption{ Dashed lines show theoretical shape of S\'{e}guy's curve centred at peak of population density $\rho=e^{-r/R}$ with $R=20$, and with time evolution accelerated by factor of 1.25. Curve computed using equation (\ref{eqn:segCity}) when $\beta=1.1, \sigma=5$ and times are $t_0 + 1.25t$ where $t=10,20,30$ and $t_0 = 4.43$. Linear scaling of time determined by maximum likelihood fit of simulation to analytical prediction. Simulation points give equivalent correlations in the full model computed from 100 independent simulations in a $400 \times 400$ system at times $t=10,20,30$.}
	\label{fig:segScaled}
\end{figure}	
Further simulations reveal that the velocity with which the city dialect expands shows some systematic deviation from the prediction $v \approx \beta \sigma^2/R$ of our OJK analysis.  For example, in Figure \ref{fig:segScaled} we have reduced the conformity parameter to $\beta=1.1$ and we see that our theoretical predictions are in close agreement with the simulation data, provided we accelerate time by a factor of $\approx 1.25$. The value of $\beta=1.4$ selected in Figure \ref{fig:seg} produces a match between predicted and observed velocity, but for larger values of $\beta$ the prediction is an overestimate. For example, when $\beta=1.5$ with all other parameters identical, the simulated velocity in the full model is smaller than our prediction by a factor of $0.97$. One possible explanation for this discrepancy is that the interface shape may effect the constant of proportionality in the Allen--Cahn equation (\ref{eqn:AC}), for example if it did not match its constant density equilibrium form. We also note that OJK's assumption of isotropy in unit normals to isoglosses, although preserved globally by the circular symmetry of our system, at the edge of the city dialect it is clearly lost locally. Despite these shortcomings the adapted OJK theory allows  analytical insight into the formation of dialects in population centres and the behaviour of S\'{e}guy's curve around cities. We leave the development of a more sophisticated theory for future work.



\section{Dialect areas and dialect continua}

There is debate amongst dialectologists as to the most appropriate way to view the geographical variation of language use \cite{cha98,hee01}. The debate arises because it is rarely the case that dialects are perfectly divided into areas. Chambers and Trudgill \cite{cha98} imagine the following example: we travel from village to village in a particular direction and notice linguistic differences (large or small) as we go. These differences accumulate so that eventually the local population are using a very different dialect from that of the village we set out from. Did we cross a border dividing the dialect area of the first village from that of the second, and if so, when? Alternatively, is it a mistake to think of dialects as organized into distinct areas; should we only think of a continuum?   

We now set out what our model can tell us about these questions. In one sense, language use in our model is always continuous in space. Although domains emerge where one variable is dominant, domain boundaries form transition regions in which the variants change continuously (the width of these regions is computed in section \ref{analysis}). Despite this, the boundary between two sufficiently large single-variant domains will appear narrow compared to the size of the domains, and in this sense the domains are well defined and noticeable by a traveller interested in one linguistic variable. Of more interest are the observations of a traveller who pays attention to the \textit{full range} of language use. To perceive a dialect boundary, this traveller must see a major change in language use over a short distance. This change must be large in comparison to other, smaller changes perceived earlier. In our model a major language change is created by crossing a large number of isoglosses over a short distance. The question then is: under what circumstances will isoglosses bundle sufficiently strongly for dialect boundaries to be noticeable?  

To answer this we need to recall the three effects which drive isogloss motion. First, surface tension which tends to reduce curvature. Second, migration of isoglosses until they emerge perpendicular to a boundary such as the coast, the border of a linguistic region, a sparsely populated zone, or an estuary. Third, repulsion of isoglosses from densely populated areas.  There are two major ways in which these effects can induce bundling, both of which require the essential ingredient of time and demographic stability in order for surface tension to take hold. Indented boundaries can collect multiple isoglosses, creating a bundle. Examples already noted include the Wash and the Severn in GB, the Gironde Estuary in France, and the historical indentation in the Dutch-German language area marking the Eastern end of the Rhenish Fan. A major boundary indentation may not always create a bundle however: it may be that other parts of the boundary, or the presence of cities, creates a fanning effect. Variations in population density can also create bundling. Dense population centres which are large in comparison to the typical interaction range will push out linguistic change, and where two centres both repel, we expect to see bundling where their zones of influence meet. Each city would then create its own well defined dialect area. Within real cities we also see sub-dialects spoken by particular social groups \cite{tru00}, but since our model does not account for social affiliations we cannot explicitly model this.

\begin{figure}
	\centering
	\includegraphics[width=\linewidth]{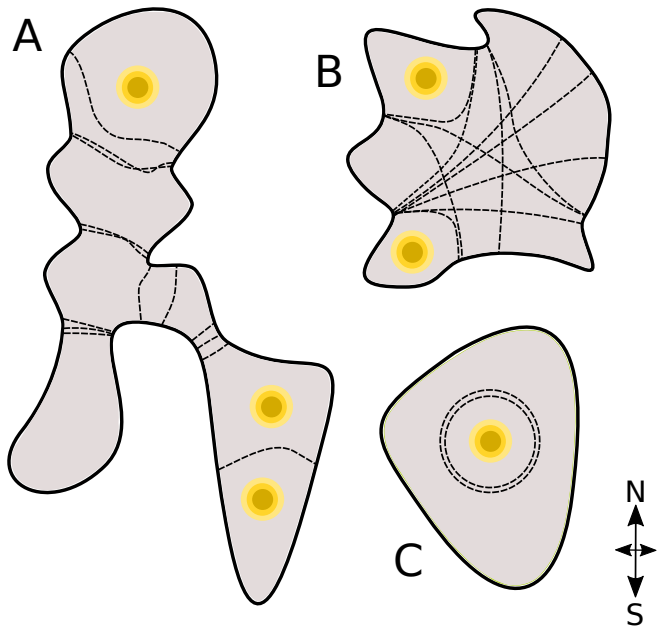}
	\caption{ Schematic diagram of isoglosses (dashed lines) for three language areas or ``Island Nations''. Yellow / ochre circles represent cities. Nation \textbf{A} supports dialect areas formed by coastal boundary shape and repulsion from cities. Nation \textbf{B} largely exhibits a continuum of variation apart from two somewhat indistinct city dialects. Nation \textbf{C} has a single city dialect, but without this city (or if the city were not sufficiently densely populated) it would have no linguistic variation due to its entirely convex boundary and evenly distributed population. }
	\label{fig:nations}
\end{figure}	
In Figure \ref{fig:nations} we schematically illustrate examples of these effects using three imaginary ``Island Nations''. Nation \textbf{A} exhibits distinct dialect areas. The Northernmost area is supported both by a city, which may generate and then repel language features, and by two indentations which form a ``pinch point'' which will \textit{tend} to collect isoglosses via the boundary effect. Several other pinches exist in which also collect isoglosses, creating distinct dialects. The Southernmost city supports an isogloss via repulsion, which would otherwise migrate south under the combined influence of surface tension and the boundary effect, eventually disappearing. Nation \textbf{B} also possesses boundary indentations, but the lack of pinch points reduces bundling: while the indentations collect isoglosses, the smoother parts of the coastline allow isoglosses to attach anywhere, creating a continuum of language use. Two city dialects do exist however, driven by repulsion. Finally, nation \textbf{C} is a convex region. This is an example of a system which, without a city, could not support more than one dialect, and would tend over time to lose isoglosses. 

In some regions there are no dialect areas, only a continuum of variation \cite{lei10}, and in others clear dialects exist \cite{tru99}. The above examples point to some general principles. In regions with low population density, a lack of major boundary indentations and few large cities, we might expect isoglosses to position themselves in a less predictable way, creating language variation which would be perceived as a continuum by a traveller. The regular creation of new isoglosses (resetting the initial conditions of the model) through linguistic innovation or demographic instability could also disrupt the ordering processes.
Narrow geographical regions, or where there are major boundary indentations, or dense population centres which push out linguistic change, are particularly susceptible to the formation of distinct dialects.

\section{Transition regions and curvature}

\label{analysis}

We now derive analytical results which characterise the transition regions between variables, and the effect of population density on the curvature of dialect domain boundaries. 

\begin{figure}
	\centering
	\includegraphics[width=\linewidth]{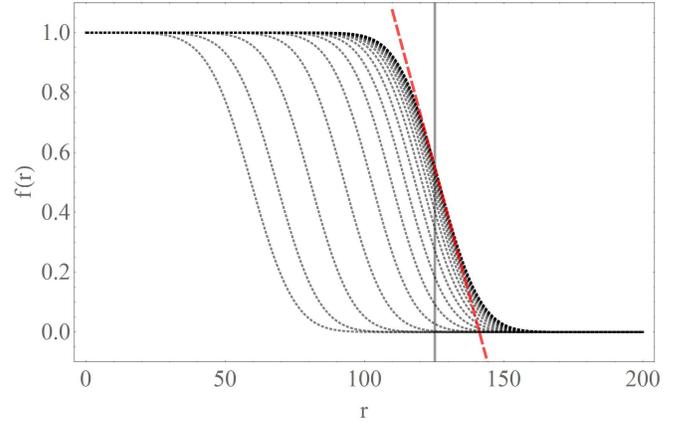}
	\caption{Sequence of radial cross-sections of the frequency of a linguistic variable whose initial domain is concentric with a city. Snapshots taken at times $t = 10,30,50,\ldots$ starting from initial radius $r=55$. Model parameters $\sigma=5, \beta=1.1$.  In this example the city has radius $R_c=50$ and peak density $\rho_1=3$. Vertical line gives theoretical stable radius $R_2=125.3$ computed from equation (\ref{stable}). Unstable radius (equation (\ref{unstable})) is $R_1=47.2$. Red dashed line gives the theoretical frequency gradient in transition region (equation (\ref{trans})).   }
	\label{fig:circ}
\end{figure}

To compute the gradients of transition regions we consider a straight isogloss (with constant population density) in equilibrium between variants 1 and 2, given by the line $x=0$. Because the isogloss is vertical the frequencies will not depend on $y$ so we write them $f_1(x)$ and $f_2(x)$ and note that $f_1(x)=1-f_2(x)$ so we need only consider the behaviour of $f_1(x)$. For notational simplicity we define $f := f_1, m:=m_1$ and $p(m):=p_1(\bv{m})$. The isogloss will form the midpoint of a transition region where the frequencies change smoothly between one and zero, and where $f_1'(0)$ measures the rate of this transition. In equilibrium, from equation (\ref{evo}) we have 
\begin{align*}
\frac{\sigma^2}{2}\p_x^2 	p(m)  = m - 	p(m).
\end{align*}
Since $f=p(m)$ we have $(\sigma^2/2) f'' = p^{-1}(f)- f$. Note that $f''$ is shorthand for $\p_x^2 f(x)$. If  non-neutrality (conformity) is small we may replace $p^{-1}(f)$ with its Taylor series about $\beta=1$, neglecting $O((\beta-1)^2)$ terms 
\begin{align}
\label{N2}
\frac{\sigma^2}{2}  f'' &= (\beta-1) f(1-f) \ln\left( \frac{1-f}{f} \right) + O((\beta-1)^2) \\
& =:-\frac{dV}{df} + O((\beta-1)^2) .
\end{align}
Here we have defined a `potential' function $V(f)$, allowing us to identify equation (\ref{N2}) as Newton's second law for the motion of a particle of mass $\sigma^2/2$ in a potential $V(f)$, where $x$ plays the role of time, so that the total `energy' $E:= \tfrac{\sigma^2}{4} f'(x)^2 + V(f(x))$ is independent of $x$ \cite{mor08}. Since $V$ is defined by an indefinite integral then we can define $V(\tfrac{1}{2}):=0$. As $x \ra \pm \infty$ we require that $f'(x)\ra 0$ and $f(x) \ra 1$ or $0$ so 
\begin{align*}
E=\lim_{f \ra 1}V(f)=\lim_{f \ra 0} V(f) = \frac{4 \ln 2-1}{24} (\beta-1). 
\end{align*}
The magnitude of the frequency gradient at the origin, where $f(0)=\tfrac{1}{2}$, is therefore	
\begin{align}
|f'(0)| =  \sqrt{\frac{4E}{\sigma^2}} \approx  0.545  \frac{\sqrt{\beta-1}}{ \sigma }.
\label{trans}
\end{align}
From this we see that weak non-neutrality, and larger interaction range produce shallower gradients and therefore wider transition regions. As $\beta$ approaches one, the transition region becomes increasingly wide and boundaries disintegrate destroying the surface tension effect described in Figure \ref{fig:tension}. Equation (\ref{trans}) is verified numerically for an `almost straight' isogloss in Figure \ref{fig:circ} (red dashed line).

We now relate the equilibrium shape of isoglosses to population density.  Starting from our modified Allen-Cahn equation (\ref{eqn:AC}) for isogloss velocity, and introducing the local radius of curvature $R = 1/\kappa$ we see that when an isogloss is in equilibrium (having zero velocity)
\begin{align}
\label{law}
R &= -\frac{\rho}{2 \nabla \rho \cdot \unit{g}},
\end{align}
where $\unit{g}$ is a unit normal to the isogloss. We note a simple alternative derivation of this result based on the \textit{dialect fraction}, $F_{\mathcal{D}}(x,y)$, of a domain $\mathcal{D}$.  We define this as the fraction of conversations that a speaker at a point $(x,y)$ has with people whose home neighbourhoods lie in $\mathcal{D}$.
\begin{align}
F_{\mathcal{D}}(x,y) &:=  \int_{\mathcal{D}} k(x,y;u,v) du dv.
\label{dfrac}
\end{align}
Let P $:=(x^\ast,y^\ast)$ be a point on an isogloss with local radius of curvature $R \gg \sigma$, bounding some region $\mathcal{D}$. An intuitively appealing condition for language equilibrium is that the speaker at P should interact with equal numbers of speakers from within and without $\mathcal{D}$ 
\begin{align} 
F_\mathcal{D}(x^\ast,y^\ast)=\tfrac{1}{2}.
\label{stableIso}
\end{align}
Using the saddle point method \cite{ma71} to evaluate equation (\ref{dfrac}) we have
\begin{align*}
F_{\mathcal{D}} &= \frac{1}{2} - \frac{\sigma}{\sqrt{2\pi}} \left(\frac{1}{2 R} +  \frac{ \nabla \rho \cdot \unit{g}}{\rho} \right) + O\left(\frac{\sigma^2}{R^2} \right).
\end{align*}
Result (\ref{law}) then follows from the equilibrium condition (\ref{stableIso}). From this we see that large gradients in population density can produce equilibrium isoglosses with higher curvature.	To test this against our full evolution equation (\ref{evo}), consider a circular city with `radius' $R_c$ having a Gaussian population distribution set in a unit uniform background 
\begin{align*}
\rho(x,y) &= 1 + (\rho_1-1) e^{-\frac{x^2+y^2}{2 R_c^2}}.
\end{align*}
The constant $\rho_1 \geq 1$ gives the relative density of the city centre compared to outlying areas. Consider a circular isogloss which is concentric with the city, then equation (\ref{law}) has two solutions
\begin{align}
\label{unstable}
R_1 &= R_c \sqrt{ \frac{1}{2} - 2 W_p\left(-\frac{e^{1/4}}{4 (\rho_1-1)} \right)} \\
R_2 &= R_c \sqrt{ \frac{1}{2} - 2 W_m\left(-\frac{e^{1/4}}{4 (\rho_1-1)} \right)} 
\label{stable}
\end{align}
where $W_p$ and $W_m$ are two branches of Lambert's W function \cite{olv10}, $W(z)$, which solves $z = w e^w$. The two solutions $R_1$ and $R_2$ are respectively decreasing and increasing functions of $\rho_1$. The stability of these solutions may be determined by noting that if $F_{\mathcal{D}}>\tfrac{1}{2}$ then $\mathcal{D}$ will expand, and contract if the inequality is reversed. From this we are able to determine that $R_2$ is stable, whereas $R_1$ is not. If a dialect domain begins with $R<R_1$ then it will shrink and disappear under the influence of surface tension, but if initially $R>R_1$ then the domain will expand or contract until $R=R_2$. This behaviour is illustrated in Figure \ref{fig:circ}, where we see that our law (\ref{law}) accurately predicts the stable radius produced by our evolution equation (\ref{evo}). 

\section{Discussion and Conclusion}

\subsection{Summary of results}

Departing from the existing approaches of dialectology, we have formulated a theory of how interactions between individual speakers control how dialect regions evolve. Much of what we have demonstrated is a consequence of the similarity between dialect formation and the physical phenomenon of phase ordering. Humans tend to set down roots, and to conform to local speech patterns. These local patterns may be viewed as analogous to local crystal orderings in binary alloys \cite{all79} or magnetic domains \cite{kra10}. As with phase ordering, surface tension is a dominant process controlling the evolution of dialect regions. However, important differences arise from the fact that human populations are not evenly distributed through space, and the geographical regions in which they live have irregular shapes. These two effects cause many different randomized early linguistic conditions to evolve toward a much smaller number of stable final states. For Great Britain we have shown that an ensemble of these final states can produce predicted dialect areas which are in remarkable agreement with the work of dialectologists.

Since language change is inherently stochastic, at small spatial scales we can only expect predictions of a statistical nature. At larger scales an element of deterministic predictability emerges.  Within our model, all stochasticity arises from the randomization of initial conditions. The effect of this randomness is strongest in the early stages of language evolution, when the typical size of dialect domains is small. The superposition of isoglosses  lacks a discernible pattern. This ``tangle'' of lines produces a continuum of language variation, with spatial correlations given by S\'{e}guy's curve. As surface tension takes hold, steered by variations in population density and system shape, isoglosses begin to bundle and the continuum resolves into distinct dialect regions. Both long term population stability, and large variations population density play an important role. Without these ingredients isoglosses will not have time to evolve into smooth lines, or bundle. 


The assumptions of our model are minimal, and clearly there are many additional complexities involved in language change which we have not captured; we discuss below how the model might be extended to account for some of these. Despite this simplicity, in addition to substantially reproducing Trudgill's predictions for English dialects \cite{tru99} we have also been able explain the observation of both dialect continua and more sharply bounded dialect domains. We have also provided an explanation for why boundary indentations (e.g. in coastline or in the border between different languages)  are likely to collect isoglosses \cite{cha98}. We have shown that cities repel isoglosses, explaining the origin of the Rhenish Fan \cite{blo00}, the wave-like spread of city language features \cite{wol03} and the fact that many dialect patterns are centred on large urban areas. We have explained why linguistic regions with high aspect ratio tend to develop striped dialect regions \cite{pre99,sam98}. We have computed an analytical form for S\'{e}guy's curve \cite{ner10,hee01,seg71,seg73} which as yet has had no theoretical derivation. We have also adapted this derivation to deal with a population centre. We have quantified how the width of a transition region \cite{cha98} between dialects is related to the strength of conformity in individual language use, and the typical geographical distances over which individuals interact. We have shown how to relate the curvature of stable isoglosses to population gradients. Finally, in appendix \ref{age} we show how incorporating an age distribution into the model can quantify the ``apparent time'' \cite{lab01} effect used by dialectologists to detect a linguistic change in progress. Given these findings, we suggest that the theoretical approach we have presented would be worth further investigation.

\subsection{Missing Pieces}

The model we have presented is deliberately minimal: it allows us to see how much of what is observed can be explained by local interactions and conformity alone. This leads to a simple unified picture with surprising explanatory power. However, having chosen simplicity, we cannot then claim to provide a complete description of the processes at play. We now describe the ``missing pieces'', indicating what effect we expect them to have, and how to include them. 

\subsubsection{Innovation}

An important aspect of language change which is not explicitly encoded within our model is \textit{innovation}: the creation of new forms of speech. We have instead assumed that there are a finite set of possible linguistic variables, and for each one, a finite set of alternatives, all of which are present in varying frequencies within the initial conditions of the model. Each alternative is equally attractive so that conformity alone decides its fate. A new variant cannot spontaneously emerge within the  domain of another. The model therefore evolves toward increasing order and spatial correlation. Due to the presence of population centres this ordering process is eventually arrested creating distinct, stable domains. If innovation were allowed, then ordering would be interrupted by the initialisation of new features, and S\'{e}guy's curve would reach a steady state where the rate of innovation (creating spatial disorder) balanced the rate of ordering. 

For a local innovation to take hold, it must overcome local conformity, realized as surface tension and the ``shrinking droplet'' effect. Several mechanisms  might allow this to happen: for example young speakers must recreate their language using input from parents, peers and other community members. This recreation process is inherently imperfect \cite{cry03}, and many interactions are between young speakers who are simultaneously assimilating their language. In this sense the young are weakly coupled to the current, adult speech community, and their language state is subject to fluctuations which may be sufficient to overcome local conformity for long enough to become established. As these speakers age their linguistic plasticity declines, older speakers die, and the change is cemented.   Other mechanisms include speech modifications made to demonstrate membership of a social group, or a bias toward easier or more attractive language features. To understand mathematically the effect of innovation on spatial evolution we might simply allow the creation of new variants, and then assign them a ``fitness'' relative to the existing population. 


\subsubsection{Interaction Network}

By selecting a Gaussian interaction kernel, and not distinguishing between different social groups, we are assuming that the social network through which language change is transmitted is only \textit{locally} connected in a geographical sense but within each locality the social network is fully connected. That is, I will listen without prejudice to anyone regardless of age, sex, status or ethnicity, as long as they live near my home. Research into 21st century human mobility \cite{bro06,gon08} and the work of linguists \cite{lab01,lab72,tru00}, indicates that both these assumptions are a simplification. Human mobility patterns, and by implication interaction kernels, exhibit heavy tailed behaviour (with an exponential cut--off at large distances).  In our framework, an interaction kernel of this type, combined with densely populated cities would allow long range connectivity between population centres. Long range interactions in phase ordering phenomena can have substantial effects on spatial correlations and domain sizes \cite{hay93}, and may effectively alter the spatial dimension of the system \cite{bla13}. 

Further evidence for non-local networks is provided by
the Frisian language, spoken in the Dutch province of Friesland. This has a ``town Frisian'' dialect \cite{hee04}, spoken in towns that are separated from each other by the Frisian countryside, where a different dialect is spoken: town Frisian is ``distributed''. Within the social network these towns are ``local'' (nearby) to each other. 
To incorporate this effect into our model we must either reformulate our fundamental equation (\ref{evo}) to describe evolution on more general network, or generalize our interaction kernel to allow communication between cities. Further empirical evidence for non-local interactions is provided by hierarchical diffusion \cite{jos03}, where linguistic innovations spread between population centres, not necessarily passing through the countryside in between. Such a process motivated the creation of Trudgill's gravity model \cite{tru74}.

In addition, mobility data \cite{gon08} shows that individuals follow regular, repeating trajectories, introducing strong heterogeneity within the set of individual interaction kernels.  Social, as well as spatial heterogeneity may also be important. For example, it has been shown theoretically \cite{pop13} that the time required for two social groups to reach linguistic consensus is highly sensitive to the level of affinity that individuals have for their own group. 

\subsubsection{Linguistic Space and Dynamics}

By using a set of \textit{linguistic variables} we are treating dialects  as points in vector space. Implicit in our dynamics are two assumptions. First, all transitions between variants are allowed, with probabilities given only by the frequencies with which the variables are used. Second, the evolution of different linguistic variables are mutually independent. There are cases where this is an incomplete description. A notable example is \textit{chain shifting} in vowel sounds \cite{lab01}. Linguists represent the set of possible vowel sounds as points in a two dimensional domain where \textit{closeness} of the tongue to the roof of the mouth, and the position of the tongue's highest point (toward the \textit{front} or \textit{back} of mouth) form two orthogonal coordinate directions \cite{roa09,dav10}. The vowel system of a language is then a set of points in this domain. If one vowel change leaves a gap in this system (an empty region of the domain) then other vowel sounds may shift to fill this gap, producing a chain of interconnected linguistic changes. Similarly, a change in one vowel to bring it closer to another can push it, and then others, out of their positions. A famous example is the \textit{Great Vowel Shift} \cite{cry03} in England between the 14th and 17th centuries.  Another example concerns changes which spread to progressively more general linguistic (as opposed to geographical) contexts \cite{mon08}. If we have several variables, each signifying the presence of the change in a different context, then it is clear that the frequency of one variable can influence the frequency of another, contradicting our assumption that variables are independent. 

The fact that linguistic variables are sometimes dependent upon one another means that, within our model, $p_i$, which relates the past use of some variable to the current frequency of its $i$th variant, via the relationship $f_i(x,y,t) := p_i[\bv{m}(x,y,t)]$, should sometimes depend on the use of other variables, and might be adapted to capture more than just conformity.

\subsection{Conclusion}

We conclude by noting that a major theme of the book \textit{War and Peace} by Leo Tolstoy, is the idea that history is determined not by \textit{great individuals} but rather by millions of small choices made by the people. 

\textit{``To elicit the laws of history we must leave aside kings,	ministers, and generals, and select for study the homogeneous,  infinitesimal  elements  which  influence  the masses.''} \cite{tol69}

\noindent
As pointed out by Vit\'{a}ny \cite{vit13}, Tolstoy was, in modern terms, advocating the formulation of a Statistical Mechanics of history. The work we have presented is an attempt to formulate such a theory for the spatial history of language. Because of its simplicity, dealing only with copying and movement, our model may apply more broadly to other forms of culture.



\appendix

\section{Discretized evolution equation}
\label{discrete}

Here we present the computational scheme used for solving our evolution equation (\ref{evo}). This is aimed at researchers having some familiarity with computer programming, such as linguists interested in quantitative approaches. It can also be implemented using only a spreadsheet (see supplementary material). The discretized version of evolution equation (\ref{evo}) also provides a greater intuitive understanding of its continuous counterpart. For simplicity we consider the $V=2$ model and define $f:=f_1$, $m:=m_1$, $p(m) := p_1(\bv{m})$ and note that we need only consider the evolution of $m$ and $f$ because $f_2=1-f_1=1-f$. 

We begin by rewriting our evolution equation (\ref{evo}) in terms of the memory and frequency fields
\begin{equation}
\label{eqn:evo2}
\p_t m = f - m + \frac{\sigma^2}{2 \rho} \nabla^2 (\rho f),
\end{equation}
where
\begin{equation}
f=p(m) = \frac{m^\beta}{m^\beta + (1-m)^\beta}
\end{equation}
To solve equation (\ref{eqn:evo2}) on a computer we discretize space into a rectangular grid of square sites. We let the side of each grid square define one unit of length.  The interaction range used in the computer calculation should be expressed in these units. That is, if the side of a grid square is $a$ km long, and the real world interaction range is $\sigma$ km, then the interaction range used in the computer should be $\sigma_c := \sigma/a$. We choose $a$ so that $\sigma_c > 1$ so speakers interact over distances greater than one square. Subscript $c$ distinguishes the interaction range measured in computer grid units from the interaction range in km. For each site we store three quantities $\rho_{ij}, f_{ij}, m_{ij}$ where the subscripts $i,j$ are horizontal and vertical indices related to spatial position by $x=i\times a$ and $y=j\times a$. These quantities are our approximations to the values of the fields $\rho,f,m$ at the centres of sites, and are stored in three arrays. Intuitively we think of each site as containing a group of $a^2 \rho_{ij}$ speakers, each of whom uses variant 1 with frequency $f_{ij}$. The linguistic domain of interest is the set of sites with non-zero population density. Within the domain, sites with one or more nearest neighbours with zero population are referred to as \textit{boundary} sites, otherwize they are \textit{bulk} sites. Sites which are not part of the domain are never updated and it is useful to include a border of such sites around the edge of the rectangular grid.

The scheme we present is based on approximating the Laplacian $\nabla^2 = \p_x^2 + \p_y^2$ using a central finite difference approximation for the derivatives $\p_x^2$ and $\p_y^2$. Let $g$ be an arbitrary function defined at every site. We define a \textit{local average} at each grid point 
\begin{equation}
\langle g \rangle_{ij} = \frac{1}{4} \left( g_{i+1,j} + g_{i-1,j} + g_{i,j+1} + g_{i,j-1} \right).
\end{equation}
This is just the average of the values of $g$ at the four nearest neighbours of $(i,j)$. The Laplacian is then approximated
\begin{equation}
\label{eqn:lapfd}
\nabla^2 g_{ij} \approx 4(\langle g \rangle_{ij} - g_{ij}).
\end{equation}
This follows from the finite difference approximation  $\p_x^2 g \approx g_{i+1,j} - 2 g_{ij} + g_{i-1,j}$, the effectiveness of which depends on $g$ varying slowly between sites.  From equation (\ref{eqn:lapfd}) we see that $\nabla^2 g$ measures the extent to which $g$ differs from the average of its neighbours. If $\nabla^2 g<0$ then $g$ exceeds the local average, and is less than the local average when the inequality is reversed.

We now introduce a small discrete time step $\dt$ and write $\Delta m_{ij} := m_{ij}(t+\dt) - m_{ij}(t)$ for the change in the memory field over the time interval $[t,t+\dt]$. We note also that provided the grid is sufficiently fine then at bulk sites  $\rho_{ij} \approx \langle \rho \rangle_{ij}$ so, making use of (\ref{eqn:evo2}) and (\ref{eqn:lapfd}) we have
\begin{align}
\label{eqn:mfd}
\Delta m_{ij} \approx \left[(f_{ij}-m_{ij}) + 2 \sigma^2 \left( \frac{\langle \rho f\rangle_{ij}}{\langle \rho \rangle_{ij}} -f_{ij} \right)\right] \dt.
\end{align}
At each time step equation (\ref{eqn:mfd}) is used to update the stored values of $m_{ij}$ for all sites in the linguistic domain, after which the frequencies can also be updated using $f_{ij} = p(m_{ij})$. The quantity $\langle \rho f\rangle_{ij}/\langle \rho \rangle_{ij}$ is the average of frequencies at neighbouring sites, weighted in proportion to their populations. Using $\langle \rho \rangle_{ij}$ rather than $\rho_{ij}$ in the denominator ensures that these weights sum to one. This serves two purposes: First, it avoids the need for an additional condition at boundary sites (intuitively, speakers in boundary sites simply shift attention from empty neighbouring squares to those which are part of the linguistic domain, consistent with the original definition of the weighted interaction kernel). Second it ensures that spatially constant memory and frequency fields constitute a fixed point of the dynamics. Rule (\ref{eqn:mfd}) is an explicit scheme and as such its stability requires that $\dt$ be chosen sufficiently small. In the case of zero conformity ($\beta=1$) and constant population density the Von-Neumann stability condition \cite{pre07} is $\dt < 1/\sigma^2$. This serves as a guide to find $\dt$ sufficiently small for our scheme to converge. For the density fields and conformity values used in this paper we have found that $\dt < 1/(4 \sigma_c^2)$ is more than sufficient. 

We conclude this section by explaining the linguistic meaning of the terms on the right hand side of (\ref{eqn:mfd}). The first term: $f_{ij}-m_{ij}$ drives conformity. If $m_{ij}>\tfrac{1}{2}$ then this term is positive, driving the memory further towards $m_{ij}=1$ where all speakers use variant 1. Otherwise if $m_{ij}<\tfrac{1}{2}$, the memory is driven towards zero where no speakers do. The second term, $\langle \rho f\rangle_{ij}/\langle \rho \rangle_{ij}-f_{ij}$, acts to equalize speech use in the local area. If variant 1 is used more at $(i,j)$ than in the surrounding squares, then this term acts to reduce its use in $(i,j)$. If variant 1 is used relatively less at $(i,j)$, the term has the opposite effect.

\section{Incorporating age into the model}
\label{age}

In order to experimentally detect a linguistic change in progress, ideally one would like to survey the same population of individuals, or a representative sample of the population, at two or more points in time \cite{cha98}. Such longitudinal studies may be practically difficult to carry out, so linguists have made use of the assumption that speech patterns are acquired mainly in the early part of people's lives. The speech of a 50-year old today should therefore reflect the speech of a 30-year old twenty years ago.  It should be noted though, that speech patterns can change throughout life \cite{san07}, although more slowly in older speakers. A linguistic change detected by observing different speech patterns in the young and old is said to have been observed in \textit{apparent time} \cite{lab63,lab01}. A famous example of apparent time is the replacement of the term \textit{chesterfield} (meaning an upholstered multiple-person seat) in Canadian speech, with the fashionable American term \textit{couch} \cite{cha95}. In this case the use of \textit{couch} was shown to decrease sigmoidally from $\approx 85\%$ amongst teenagers to $\approx 5\%$ among those in their eighties. The apparent time theory has been tested by comparing language surveys taken at different times, and comparing predictions based on apparent time in the earlier sample, with the observations made in the later one \cite{bai91}.  We note that differences between speech patterns between the generations do not always indicate a linguistic change in progress \cite{san07}. For example, the use of some speech forms may change systematically with age in the same way, generation after generation, so that the community as a whole is in a stable state \cite{tru00}.

\begin{figure}
	\centering
	\includegraphics[width=\linewidth]{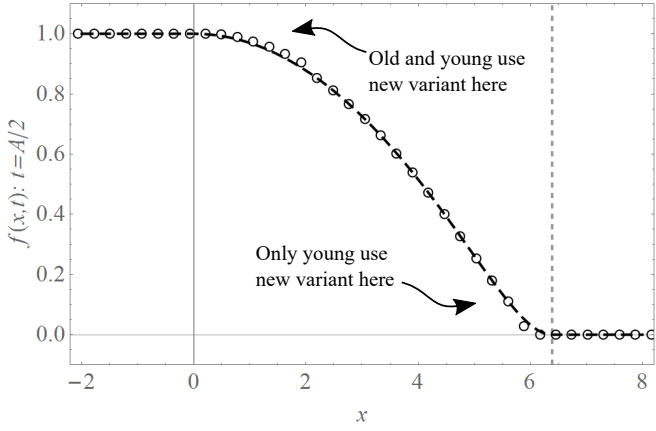}
	\caption{ Thick dashed line shows theoretical frequency of new variant for age distribution (\ref{eqn:age}) -- using linear approximations (\ref{eqn:v}) and $(\ref{eqn:Delta})$ for $v(a)$ and  $\Lambda(a)$, and assuming $\Lambda(0)=0$ -- when $a_0=10, A=90, \sigma=5,\epsilon=0.1$ at time $t=A/2=45$ (chosen so that oldest speaker at $x=0$ has just switched variable). For these parameter values $\bar{a}=35.23$.  Vertical dashed line is drawn at location of youngest adopter of new variable (giving width of transition region). Open circles give values of boundary using an age distribution discretized into two year bins, computed by numerically evolving the full model.   }
	\label{fig:memwav}
\end{figure}

We now give a simple illustration of how age and apparent time can be incorporated in our model. For simplicity we consider the progress of a straight isogloss between two variants, driven by a slowly declining population density. This density variation is equivalent to a social bias toward one variable which we call the \textit{new variant}. We let $f(x,t)$ be the frequency with which the new (spreading) variable is used at position $x$ and time $t$. Note that there is no $y$ dependence due to symmetry.
To distinguish between young and old we introduce an age density distribution $\alpha(a)$ giving the fraction of individuals within the age bracket $[a_1,a_2]$ as
\begin{equation}
\int_{a_1}^{a_2} \alpha(a)da.
\end{equation}
Using this distribution we modify our original model to account for the fact that 
individuals have been exposed only to the linguistic information available in their lifetime. The memory of a speaker with age $a$ is therefore defined
\begin{equation}
\mu_a(x,t) := \int_0^a \frac{e^{-\frac{s}{\tau}}}{\tau(1-e^{-a/\tau})} \left[  \int_{\mathbb{R}}  k(x;u)f(u,t-s) du \right] ds.
\end{equation}
Note that as $a \ra \infty$ this definition coincides with our original definition (\ref{eq:m}) of memory. In the interests of analytical tractability we consider the limit of small memory decay rate ($\tau \ra \infty$) in which case linguistic memory is a simple ``bus stop'' average over life history
\begin{equation}
\mu_a(x,t) = \frac{1}{a} \int_0^a \left[ \int_{\mathbb{R}} k(x;u) f(x,t-s) du \right] ds.
\end{equation}
We also take the limit of total conformity $\beta \ra \infty$ so that language is chosen according to a simple majority rule.  We consider an exponentially decaying population density
\begin{equation}
\rho(x) = e^{-\epsilon x}
\end{equation}
where $\epsilon \ll 1$. The weighted interaction kernel for this density is then
\begin{align}
k(x;u) &= \frac{1}{\sqrt{2\pi} \sigma} \exp \left\{ -\frac{(u-x+\epsilon \sigma^2)^2}{2\sigma^2}\right\} \\
&=: k(u-x).
\end{align}
Notice that the effect of the decaying population density is to shift the interaction kernel to the left so that more attention is paid to language use on that side of the listener. To compute the isogloss velocity we define 
\begin{equation}
\eta(x) := \begin{cases}
1 &\text{ if } x<0 \\
\frac{1}{2} &\text{ if } x=0 \\
0 &\text{ if } x>0 
\end{cases}
\end{equation}
and prepare the system with initial condition
\begin{equation}
f(x,t <t_0) = \eta(x-\Lambda)
\end{equation}
where $\Lambda$ is the initial location of the isogloss. Because each speaker listens more to the speakers on her left, the isogloss will travel right. In the limit $\beta \ra \infty$ then when the memory, $\mu_a(x,t)$, of a speaker, with $x > \Lambda$, reaches $\tfrac{1}{2}$ they will switch linguistic variables.
The motion of the isogloss will then take the form of a travelling wave formed from a superposition of travelling step functions, one for each age
\begin{equation}
f(x,t) = \int_0^\infty \alpha(a) \eta(x-vt+\Lambda(a)) da,
\label{eqn:wav}
\end{equation}
with the function $\Lambda(a) > 0$ and the velocity $v$ to be determined. According to (\ref{eqn:wav}), at $t_a:=\Lambda(a)/v$, a speaker at the origin with memory of length $a$ will be on the point of switching variable so that
\begin{multline}
\mu_a(0,t_a) = \frac{1}{2} =\\
 \int_0^a \frac{ds}{a}  \int_0^\infty da' \int_{\mathbb{R}} dy \  
\alpha(a') k(y) \eta[y + v s -\Lambda(a) + \Lambda(a')]  \\
= \int_0^a \frac{ds}{a}  \int_0^\infty da' \alpha(a') \int_{-\infty}^{\Lambda(a)-\Lambda(a')-vs} k(y) dy \\
:= \int_0^a \frac{ds}{a}  \int_0^\infty  \alpha(a') K(\Lambda(a)-\Lambda(a')-vs) da'.
\label{eqn:cond}
\end{multline}
Here we have introduced the cumulative $K$ of the interaction kernel
\begin{equation}
K(z) := \int_{-\infty}^z k(y) dy.
\end{equation}
As $\epsilon \ra 0$ the isogloss velocity must also tend to zero. The quantity
\begin{equation}
\Delta(a_1,a_2) := \Lambda(a_1)-\Lambda(a_2)
\end{equation}
gives the distance between the step functions for speakers with ages $a_1$ and $a_2$ as $|\Delta(a_1,a_2)|$, and this separation must also tend to zero as $\epsilon \ra 0$. We can therefore compute a series expansion for $v$ in powers of $\epsilon$ by expanding the cumulative interaction kernel $K(z)$ in (\ref{eqn:cond}) about $z=0$ and $\epsilon=0$. To lowest order we have
\begin{equation}
K(z) = \frac{1}{2} + \frac{z}{\sqrt{2\pi} \sigma} + \frac{\epsilon \sigma}{\sqrt{2\pi}} + o(z) + o(\epsilon).
\end{equation}
Substituting this approximation into equation (\ref{eqn:cond}) we have
\begin{equation}
\int_0^\infty \frac{\Delta(a,a')}{\sigma} \alpha(a') da' - \frac{av}{2 \sigma} + \epsilon \sigma = 0.
\end{equation}
It is straightforward to verify that this equation has solution
\begin{align}
\label{eqn:v}
v &= \frac{2 \sigma^2 \epsilon}{\bar{a}} \\ 
\Delta(a,a') &= \frac{(a-a') \sigma^2 \epsilon}{\bar{a}} 
\label{eqn:Delta}
\end{align}
where $\bar{a}$ is the mean age of the population
\begin{equation}
\bar{a} := \int_0^\infty a \alpha(a) da.
\end{equation}
If the oldest speaker has age $A$, then the width of the transition region is $\Delta(A,0) = A \sigma^2 \epsilon/\bar{a}$. We provide a concrete example using a population ``pyramid'' age distribution, cut off exponentially at low ages to account for the fact that very young speakers listen to, but do not influence, others. Letting $a_0$ be the low age cut-off we define
\begin{equation}
\alpha(a) = \frac{1}{\mathcal{C}} (1-e^{-a/a_0})(A-a),
\label{eqn:age}
\end{equation}
where $\mathcal{C}$ is a normalizing constant. An example of the travelling wave (\ref{eqn:wav}) generated by this age distribution is illustrated in Figure \ref{fig:memwav}. Also shown are the results of a numerical implementation of the full model with a discretized version of the age distribution (\ref{eqn:age}). This discretization is necessary in order to implement the model numerically, because the memory of each age of speaker must be individually stored.

Finally we consider the likely outcome of experimentally sampling the use of language within this model. We let $x_0$ be the left boundary of the transition region (the oldest speaker at $x_0$ is just about to switch variable).  We then define the indicator function of the event that a speaker of age $a$, located at position $x$, is using the new variant
\begin{equation}
q_x(a) := \begin{cases}
1 &\text{ if } x-x_0 < \Delta(A,a) \\
0 & \text{ otherwize. }
\end{cases}
\end{equation}
Consider a sample of speakers with home locations, $X$, normally distributed around some average position $x_0+h$: $X \sim \mathcal{N}(x_0+h, \omega^2)$. The probability that a speaker of age $a$ within this sample will use the new variant is then expectation of $q_X(a)$ over the position $X$
\begin{equation}
 \EE[q_X(a)] = \frac{1}{2} \text{erfc} \left( \frac{h-\Delta(A,a)}{\sqrt{2} \omega}\right)  =: \bar{q}(a;h,\omega).
 \label{eqn:ageDist}
\end{equation}
\begin{figure}
	\centering
	\includegraphics[width=\linewidth]{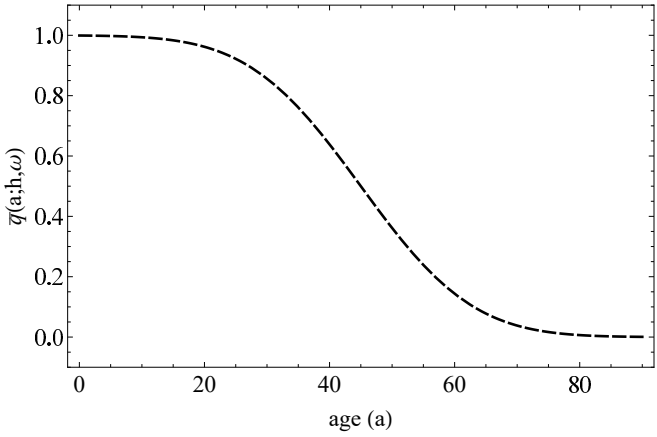}
	\caption{ Expected frequency (\ref{eqn:ageDist}) with which new variant is used by speakers of different ages from a random sample. Mean and variance of speaker locations relative to left boundary $x_0$ of transition region are $h=\tfrac{1}{2} \Delta(0,A)$ and $\omega^2=1$. Model parameter values are $a_0=10, A=90, \sigma=5,\epsilon=0.1$. For these parameter values $\bar{a}=35.23$.    }
	\label{fig:ageDist}
\end{figure}
An example of this distribution is illustrated in Figure \ref{fig:ageDist}. In this example the mean sample location is the centre of the transition region, and we see that uptake of the new variant exhibits the characteristic ``S-shaped'' age distribution seen in studies of linguistic change observed in ``apparent time'' \cite{cha95,lab01,cha98}.	

We conclude by noting that this extension of the model to include different memory lengths should be seen as a ``toy'' model of the effect of age on language change. The fact that older speakers tend to take longer to change their language use is captured purely by the length of their memories. In reality the influence on a given speaker of the language she is exposed to at different stages of her life will be much more complicated than our simple model \cite{bax16}. For example, if language use were determined entirely during early life then the \textit{forgetting curve} should be peaked during these early years - in effect the linguistic memory should stop ``recording'' once a speaker's youth has ended. Each speaker will respond differently to what they hear, so the forgetting curve will not be identical for every speaker. Such heterogeneity amongst speakers is straightforward to incorporate, but at the cost of tractability. The advantage of the simple approach is to illustrate the apparent time effect in an analytically simple way. 

\section{Allen--Cahn Equation and Ohta--Jasnow--Kawasaki Theory} 
\label{AC}
Here we derive an analogue of the Allen--Cahn equation \cite{all79} for the velocity of an isogloss, before deriving a modified Ohta--Jasnow--Kawasaki equation \cite{oht82} which provides a simplified method for understanding the evolution of spatial correlations in the model. In the binary variant case ($V=2$) we have that $m_1=1-m_2$ so after defining $p(m_1):=p_1(\bv{m})$ and $f:=f_1$ then $f=p(m_1)$. In terms of $f$, our evolution equation (\ref{evo}) may be written 
\begin{equation}
\p_t f = p'\left[ p^{-1}(f) \right]\left\{f-p^{-1}(f) + \frac{\sigma^2}{2\rho} \nabla^2 (\rho f) \right\}. 
\label{eqn:f}
\end{equation}
Following \cite{all79,bra94} we introduce a unit vector, $\unit{g}$, normal to the isogloss at a point $P$. We let $g$ be the displacement from $P$ in the direction of $\unit{g}$. At the isogloss we have
\begin{align*}
\nabla f &= \left(\frac{\p f}{\p g}\right)_t \unit{g} \\
\nabla^2 f &= \left(\frac{\p^2 f}{\p g^2}\right)_t + \left(\frac{\p f}{\p g}\right)_t \nabla \cdot \unit{g}. 
\end{align*}
We will also need the cyclic relation
\begin{equation}
\left(\frac{\p t}{\p f}\right)_g \left(\frac{\p f}{\p g}\right)_t \left(\frac{\p g}{\p t}\right)_f= -1 .
\label{cr}
\end{equation}
The laplacian term in equation \ref{eqn:f} may be expanded as follows
\begin{align}
 \frac{\nabla^2 (\rho f)}{\rho} &=  \nabla^2 f + f \frac{\nabla^2{\rho}}{\rho} + 2 \frac{\nabla \rho \cdot \nabla f}{\rho} \\
 &= \left(\frac{\p^2 f}{\p g^2}\right)_t + \left(\frac{\p f}{\p g}\right)_t \nabla \cdot \unit{g}  + f \frac{\nabla^2{\rho}}{\rho} + 2 \frac{\nabla \rho \cdot \nabla f}{\rho}
 \label{eqn:lap}
\end{align}
For constant density, the equilibrium configuration of the isogloss is a straight line so the curvature $\kappa := \nabla \cdot \unit{g}$ is zero, and $f$ depends only on the displacement $g$. Making use of (\ref{eqn:f}) and (\ref{eqn:lap}) we see that the equilibrium equation for $f$ is in this case
\begin{equation}
\frac{\sigma^2}{2} \frac{d^2f}{dg^2} = p^{-1}(f)-f.
\end{equation}
We now make the assumption that out of equilibrium, if the curvature is low, and the density slowly varying, then the profile of the transition region around the isogloss takes its equilibrium form. We also recall our assumption in our derivation of the full evolution equation (\ref{evo}) that $|\nabla^2 \rho|/\rho \ll \sigma^2$. The evolution equation (\ref{eqn:f}) may then be written
\begin{align}
\p_t f &= \frac{\sigma^2}{2} p'\left[ p^{-1}(f) \right]\left\{  \left(\frac{\p f}{\p g}\right)_t \nabla \cdot \unit{g}  + 2 \frac{\nabla \rho \cdot \nabla f}{\rho}\right\} \\
&=  \sigma^2 p'\left[ p^{-1}(f) \right] \left(\frac{\p f}{\p g}\right)_t\left\{   \frac{\nabla \cdot \unit{g}}{2}  + \frac{\nabla \rho \cdot \unit{g}}{\rho}\right\}
\end{align}
Making use of relation (\ref{cr}) we have
\begin{equation}
\left(\frac{\p g}{\p t} \right)_f= -\sigma^2 p'\left[ p^{-1}(f) \right] \left\{   \frac{\kappa}{2}  + \frac{\nabla \rho \cdot \unit{g}}{\rho}\right\}
\end{equation}
Since $(\p g/\p t)_f$ is the isogloss velocity, and at the isogloss we have $f=\tfrac{1}{2}$ then
\begin{equation}
v = -\sigma^2 \beta \left\{   \frac{\kappa}{2}  + \frac{\nabla \rho \cdot \unit{g}}{\rho}\right\}.
\end{equation}
To obtain spatial correlation functions between different modal linguistic variables we apply the Ohta-Jasnow-Kawasaki (OJK) method \cite{oht82}. We adapt the description of OJK's analysis given in Bray \cite{bra94} to include population density effects.  As described in section \ref{seg} we label the two alternatives for a particular variable as $-1$ and $1$ and  introduce a smoothly varying auxiliary field $m(x,y,t)$ which gives the modal (most common) variant $\phi_i$, of variable $i$ as $\phi_i(m) = \sgn (m)$.
The unit vector $\unit{g}$ may then be written
\begin{equation}
\unit{g} = \frac{\nabla m}{|\nabla m|},
\end{equation}
from which we see that the isogloss velocity is
\begin{equation}
v = \sigma^2 \beta \left(\frac{-\nabla^2 m + \sum_{i,j} \hat{g}_i \hat{g}_j \p_i \p_j m - 2(\nabla \rho.\nabla m)/\rho }{2|\nabla m|} \right),
\end{equation}
where $i,j \in \{x,y\}$. In a reference frame attached to the interface
\begin{equation}
\frac{dm}{dt} = 0 = \frac{\p m}{\p t} + \bv{v}\cdot \nabla m.
\end{equation}
Since $\bv{v} \parallel \nabla m$ then $\bv{v}\cdot \nabla m = v |\nabla m|$ and $\p_tm = - v |\nabla m|$ so
\begin{equation}
\frac{\p m}{\p t} =   \frac{\sigma^2 \beta}{2} \left(\nabla^2 m - \sum_{i,j} \hat{g}_i \hat{g}_j \p_i \p_j m + 2 \frac{\nabla \rho.\nabla m}{\rho} \right).
\end{equation}
This is the OJK equation, modified to include variable population density. As OJK did, we now assume that the direction $\unit{g}$ is uniformly distributed  over the system, and we replace $g_i g_j$ with its circular mean $\tfrac{1}{2} \delta_{ij}$, giving
\begin{equation}
\frac{\p m}{\p t} =  \sigma^2 \beta \left(\frac{\nabla^2 m}{4} + \frac{\nabla \rho.\nabla m}{\rho} \right)
\end{equation}
This is our modified Ohta--Jasnow--Kawasaki equation.

\section{Comparison with a simulation of Nerbonne}

\label{ner}

We note a link between curve (\ref{eqn:seg}) and a simple model simulated by Nerbonne \cite{ner10}, but not characterized analytically. The model consists of a line of discrete spatial points, with a single reference site at one end representing a city. In contrast with our model, all sites initially use an identical set of $N$ ($=100$ in \cite{ner10}) binary linguistic variables. Evolution of language use is simulated at each site by repeatedly selecting a variable at random and then changing the state of the variable with probability $\tfrac{1}{2}$. For a site at distance $r$ from the city, the number of repeats of this randomization process is defined $n(r) := \lfloor C r^\alpha \rfloor$, where the constant $\alpha$ measures the spatial decline of the ``influence'' of the city with $r$. Larger values of $n(r)$ imply a greater level of noisy evolution and therefore a lower influence of the city. Linguistic distance in this model, after each site has received its $n(r)$ updates, is given by
\begin{equation}
l(r) = \tfrac{1}{2} \left[1- (1-N^{-1})^{n(r)}\right] \approx \tfrac{1}{2} \left[ 1- e^{-\frac{n(r)}{N}} \right].
\end{equation}
In \cite{ner10} two values of $\alpha$ are tested: $\alpha=1,2$, and the quadratic case is identified as being consistent with Trudgill's macroscopic gravity model. However, we emphasize that the two models do not make predictions that can be directly compared: in the microscopic model \cite{ner10} no indication is given of how two centres of influence would compete. Of interest is the fact that the $\alpha=2$ case coincides with our prediction for large $r$. However, this value of $\alpha$ is rejected in \cite{ner10} on the basis of its sigmoidal shape for small $r$.  Due to the presence of the inverse sine function this behaviour is not present in our version of S\'{e}guy's law (\ref{eqn:seg}). We note also that the dynamics of \cite{ner10} reduce order in the system; whereas our model leads to increasingly ordered states.

\begin{acknowledgments}
The author was supported by a Leverhulme trust research fellowship while carrying out this work, and is most grateful for this. The referees, from Statistical Physics and Dialectology, provided invaluable advice and their time is greatly appreciated. The author would like to particularly acknowledge one quantitative dialectologist reviewer whose comments had a substantial impact on the work.  He is also particularly grateful to Samia Burridge for many ideas and useful advice.
\end{acknowledgments}

\bibliography{}
\end{document}